\documentclass[%
reprint,
 amsmath,amssymb,
aps,
]{revtex4-2}

\usepackage{graphicx}
\usepackage{dcolumn}
\usepackage{bm}
\usepackage{hyperref}
\usepackage{xcolor}
\usepackage[normalem]{ulem}

\newcommand{\upperRomannumeral}[1]{\uppercase\expandafter{\romannumeral#1}}

\begin{document}

\preprint{APS/123-QED}

\title{Kinematic signatures of impulsive supernova feedback in dwarf galaxies}

\author{Jan D. Burger}
\email{burger@mpa-garching.mpg.de}
\affiliation{University of Iceland, Dunhagi 5, 107 Reykjavík, Iceland}
\affiliation{Department of Physics and Astronomy, University of California, Riverside, CA, 92521, USA}
\affiliation{Max-Planck-Institut f\"ur Astrophysik, Karl-Schwarzschild-Str. 1, 85748 Garching, Germany}%
\author{Jesús Zavala}
\affiliation{University of Iceland, Dunhagi 5, 107 Reykjavík, Iceland}%
\author{Laura V. Sales}
\affiliation{Department of Physics and Astronomy, University of California, Riverside, CA, 92521, USA}%
\author{Mark Vogelsberger}
\affiliation{MIT Kavli Institute for Astrophysics and Space Research, Ronald McNair Building, 37-611}%
\affiliation{The NSF AI Institute for Artificial Intelligence and Fundamental Interactions, Massachusetts Institute of Technology, 77 Massachusetts Ave, Cambridge MA 02139, USA}
\author{Federico Marinacci}
\affiliation{Department of Physics and Astronomy ``Augusto Righi'', University of Bologna, Bologna, I-40129, Italy}%
\author{Paul Torrey}
\affiliation{Department of Astronomy, University of Florida, 211 Bryant Space Sciences Center, Gainesville, FL 32611, USA}%




\date{\today}

\begin{abstract}
Impulsive supernova feedback and non-standard dark matter models, such as self-interacting dark matter (SIDM), are the two main contenders for the role of the dominant core formation mechanism at the dwarf galaxy scale.
Here we show that the impulsive supernova cycles that follow episodes of bursty star formation leave distinct features in the distribution function of stars: groups of stars with similar ages and metallicities develop overdense shells in phase space. If cores are formed through supernova feedback, we predict the presence of such features in star-forming dwarf galaxies with cored host halos. Their systematic absence would favor alternative dark matter models, such as SIDM, as the dominant core formation mechanism.
\end{abstract}

\maketitle

\begin{figure}
    \centering
    \includegraphics[width=\linewidth, trim={0.5cm 0.5cm 0.5cm 0.5cm},clip=true]{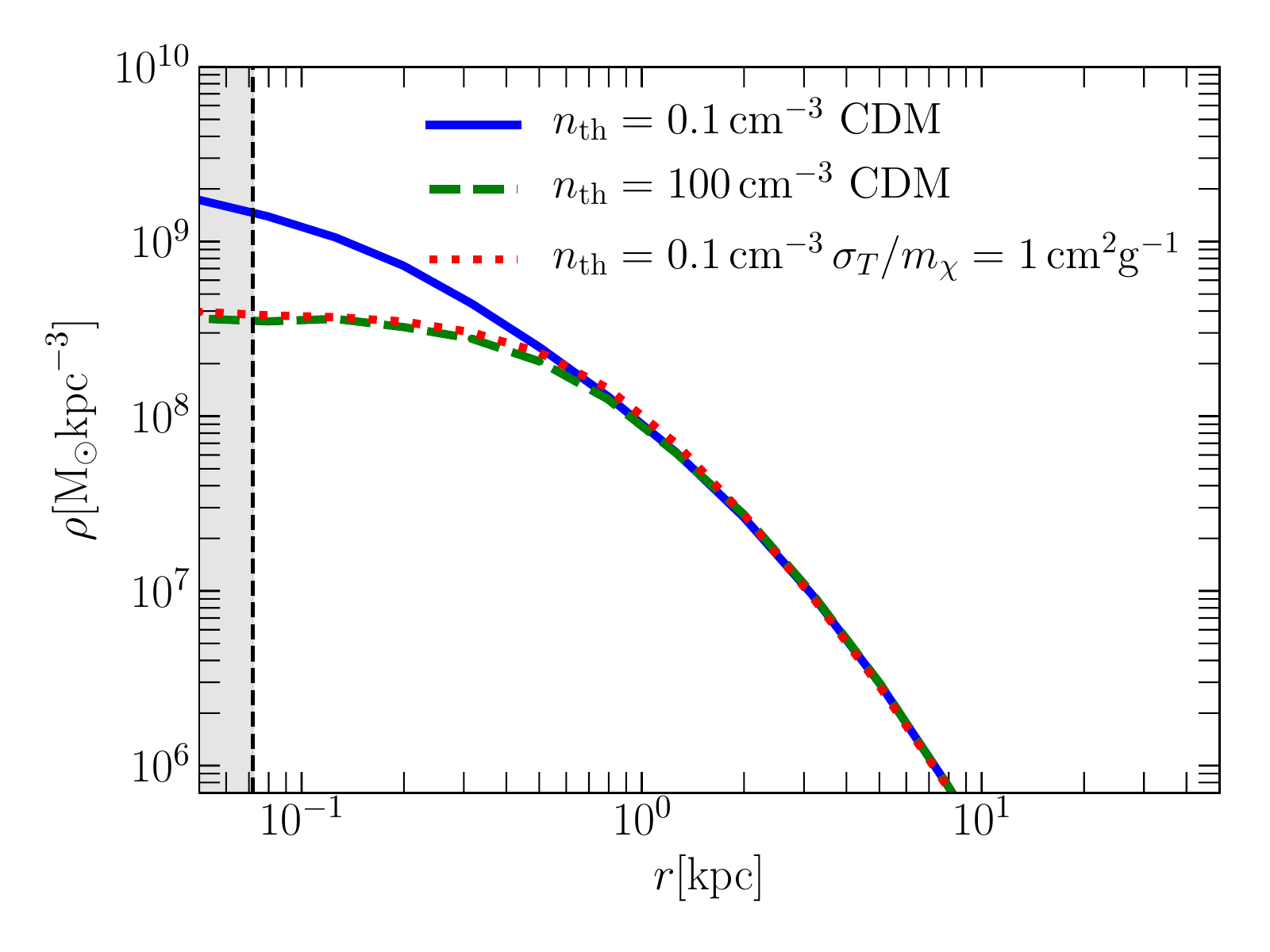}\\
    \caption{Adiabatic and impulsive cusp-core transformation: Spherically averaged radial DM density
    of three different simulated dwarf-size halos after $3\,$Gyr of simulation time. The solid blue line corresponds to the CDM simulation with smooth star formation ($n_{\rm th} = 0.1\,{\rm cm^{-3}}$), whereas the dashed green line denotes the result of the CDM simulation with bursty star formation ($n_{\rm th} = 100\,{\rm cm^{-3}}$). 
    The red dotted line corresponds to the SIDM simulation with smooth star formation ($n_{\rm th} = 0.1\,{\rm cm^{-3}}$) 
    and a self-interaction cross section $\sigma_T/m_\chi = 1\,{\rm cm^2g^{-1}}$. 
    }
    \label{fig:dm_profiles}
\end{figure}

\begin{figure}
    \centering
    \includegraphics[width=0.48\linewidth]{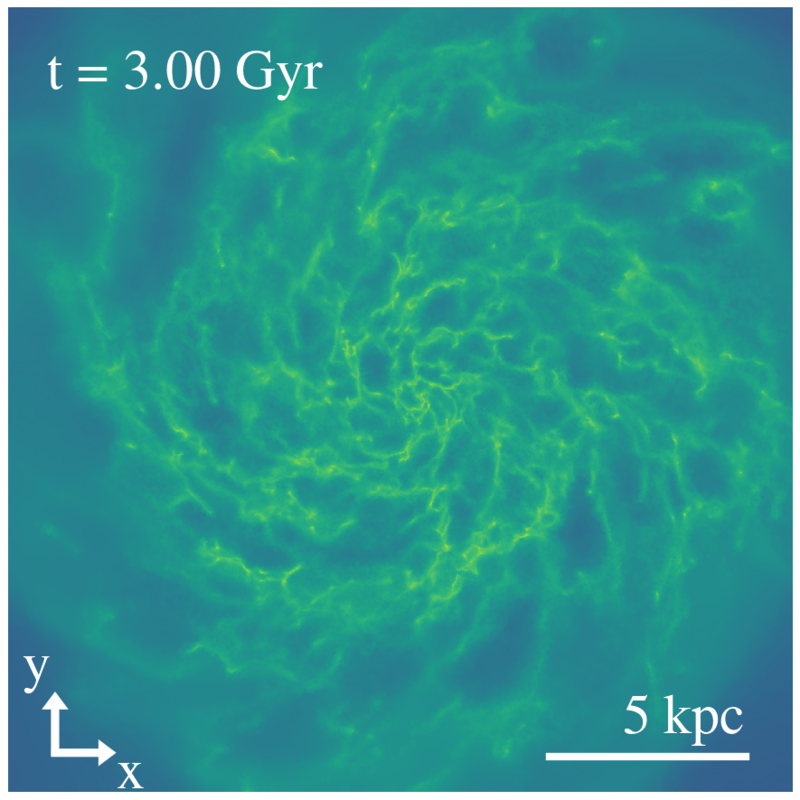}
    \includegraphics[width=0.48\linewidth]{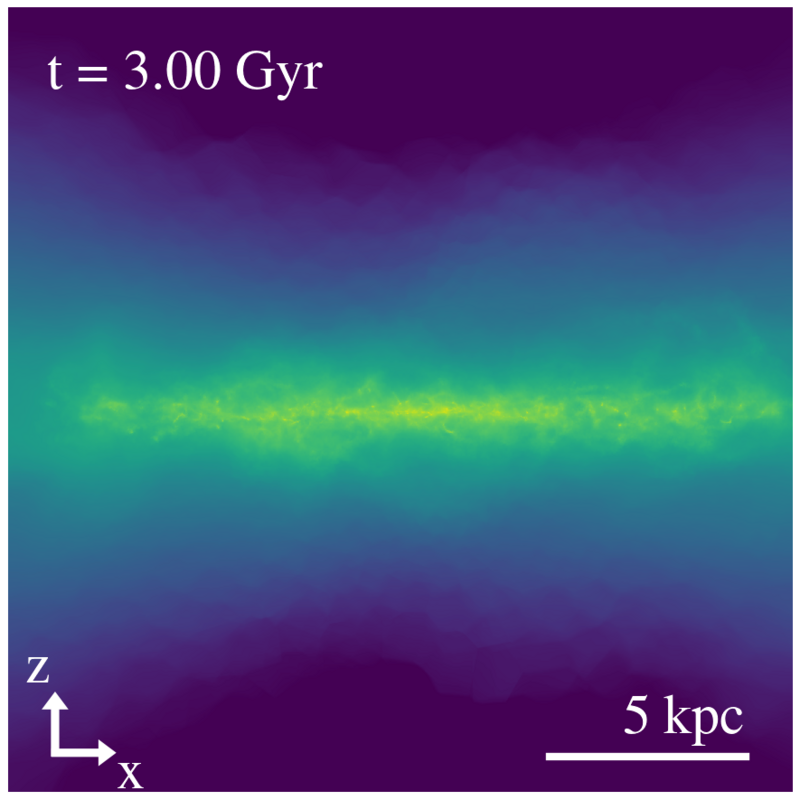}\\
    \includegraphics[width=0.48\linewidth]{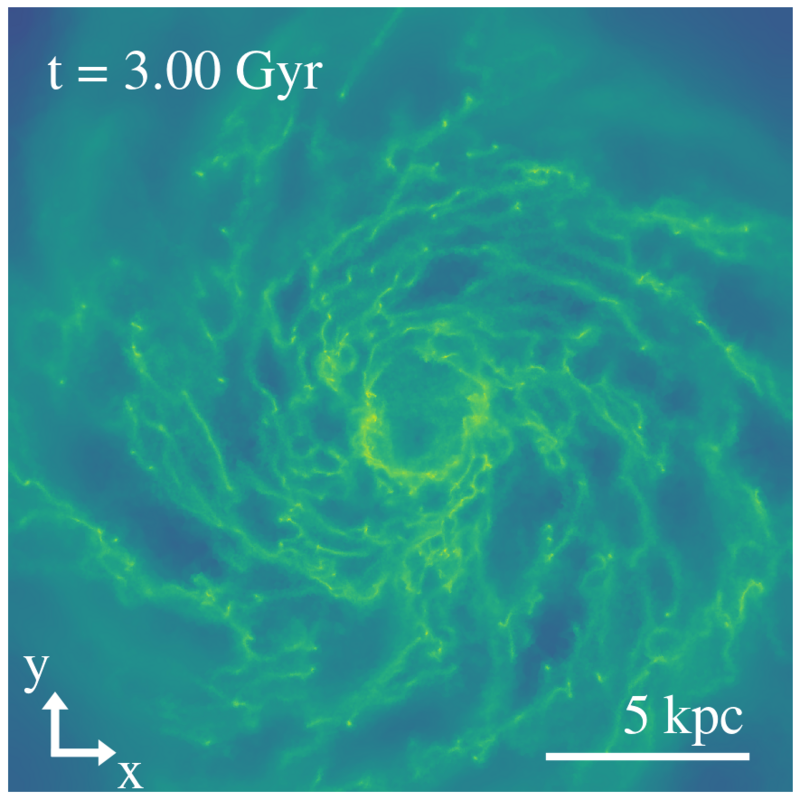}
    \includegraphics[width=0.48\linewidth]{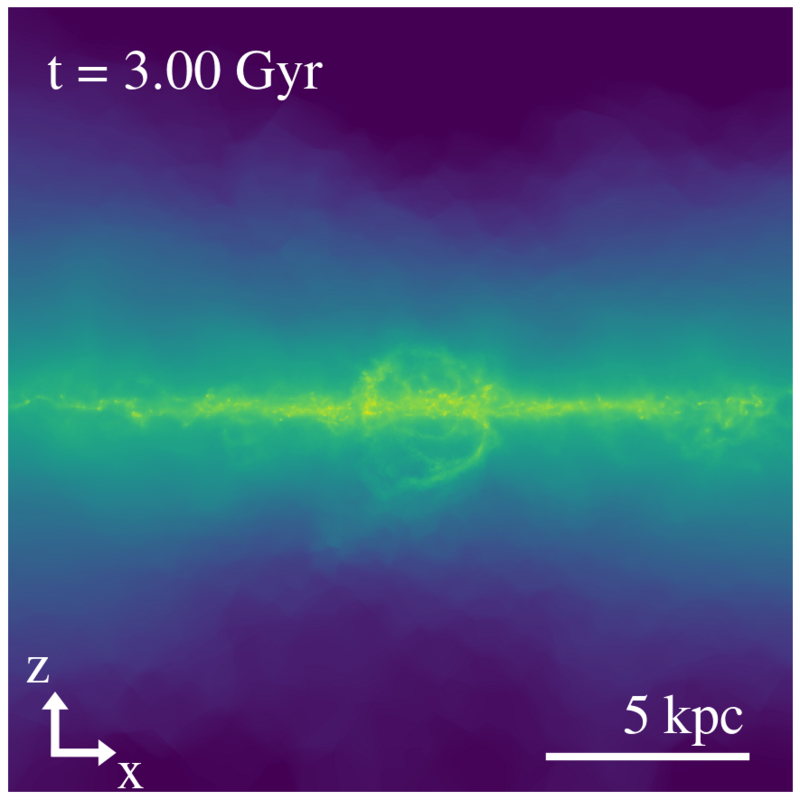}

    \caption{Gas distributions of resulting galaxies after $3\,$Gyr of simulation time for two simulations with different star formation thresholds. 
    We show face-on (left column) and edge-on (right column) projections of the gas density
    for the CDM simulation with smooth (top row) and bursty (bottom row) star formation. The side-length of the field of view is $20\,$kpc in each panel and we defined a coordinate system such that the z-axis is perpendicular to the gas disc. 
    Notice how central gas is vertically expelled out of the disc plane in the simulation with bursty star formation. Such galactic outflows are characteristic of violent and impulsive events of supernova-driven energy release.} 
    \label{fig:galaxyview}
\end{figure}

\begin{figure*}
    \centering
    \includegraphics[width=8cm,trim={0.6cm 0.6cm 0.5cm 0.5cm},clip=true]{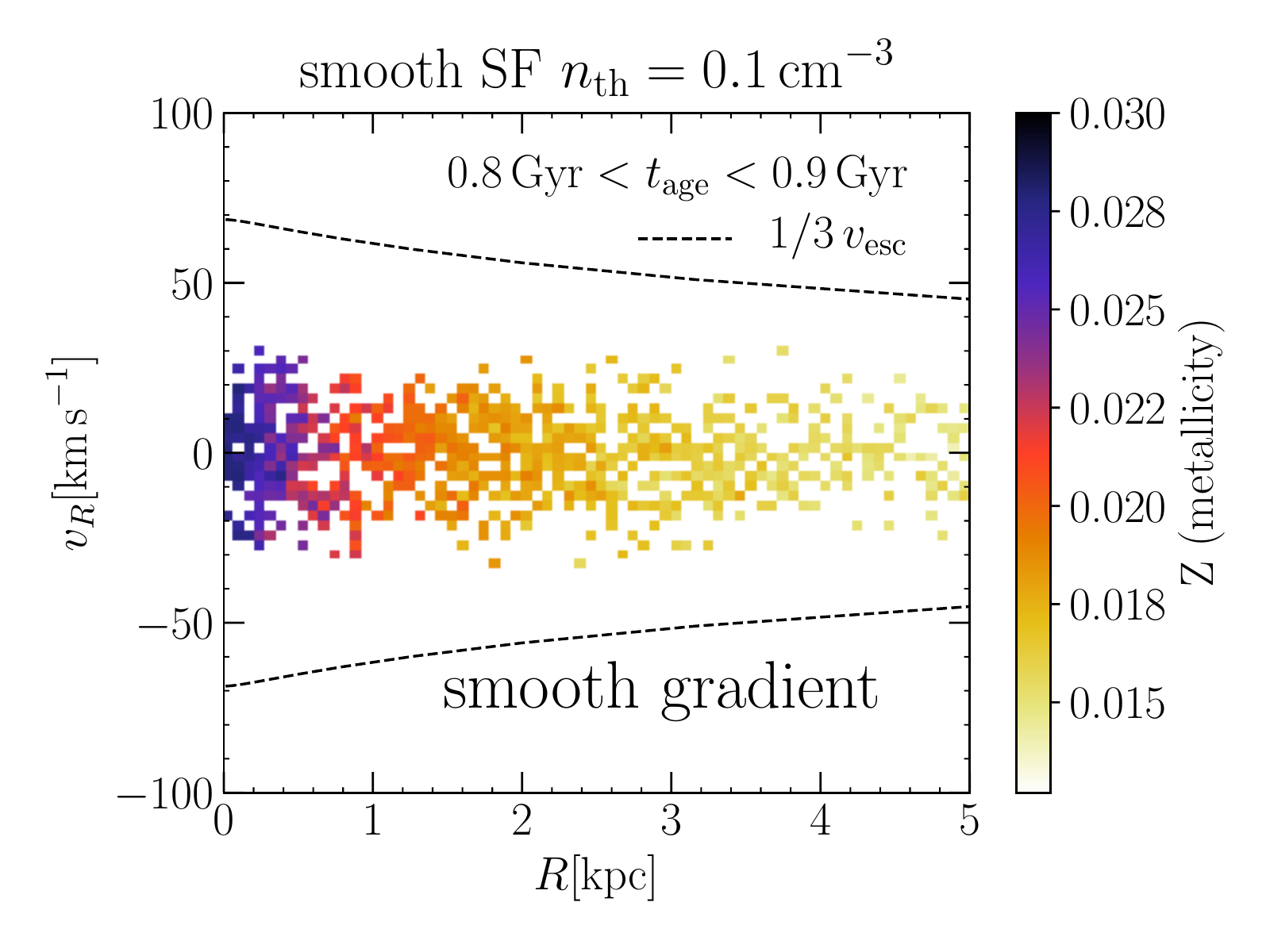}
    \includegraphics[width=8cm,trim={0.6cm 0.6cm 0.5cm 0.5cm},clip=true]{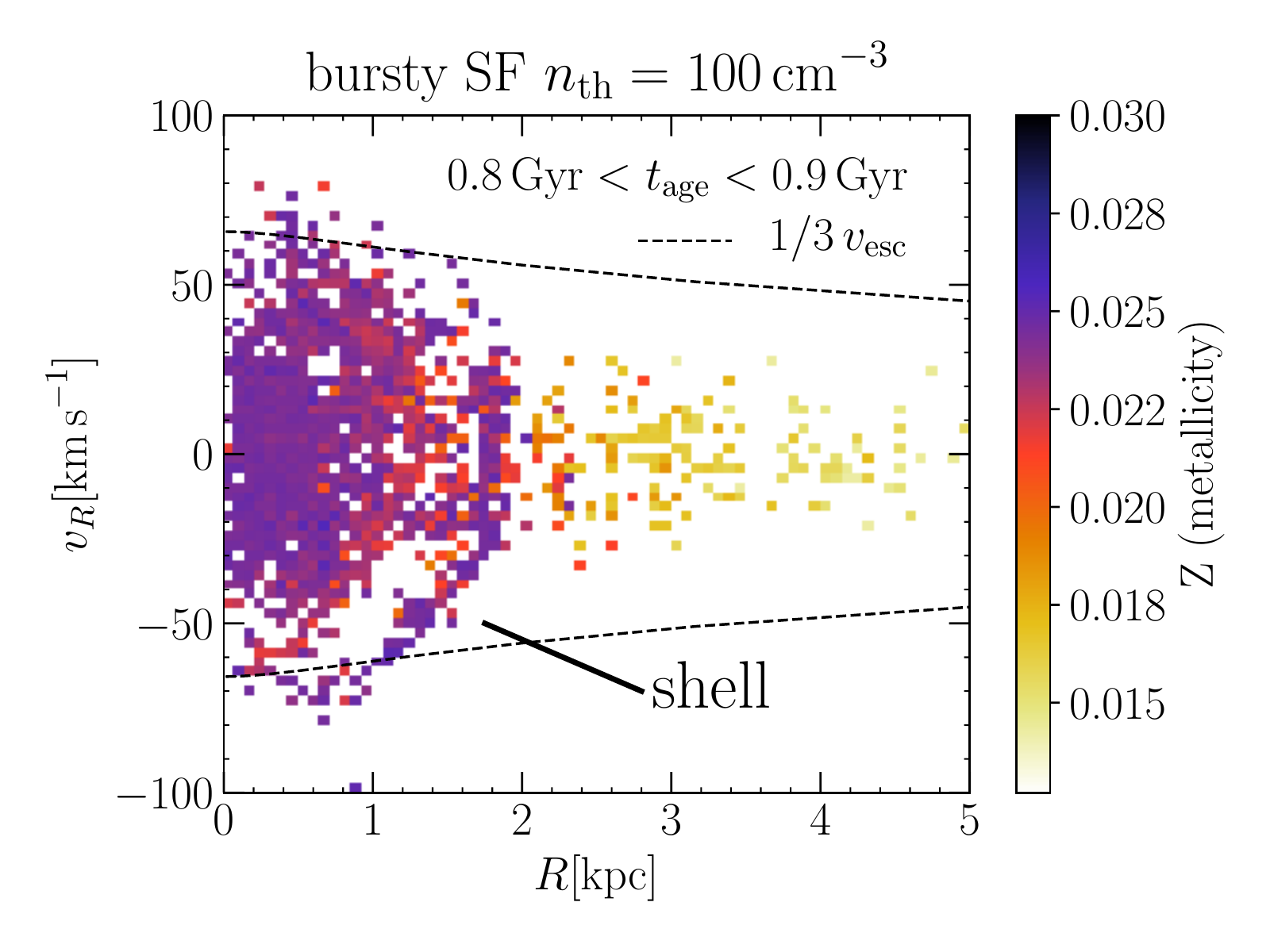}\\
    \includegraphics[width=8cm,trim={0.6cm 0.6cm 0.5cm 1cm},clip=true]{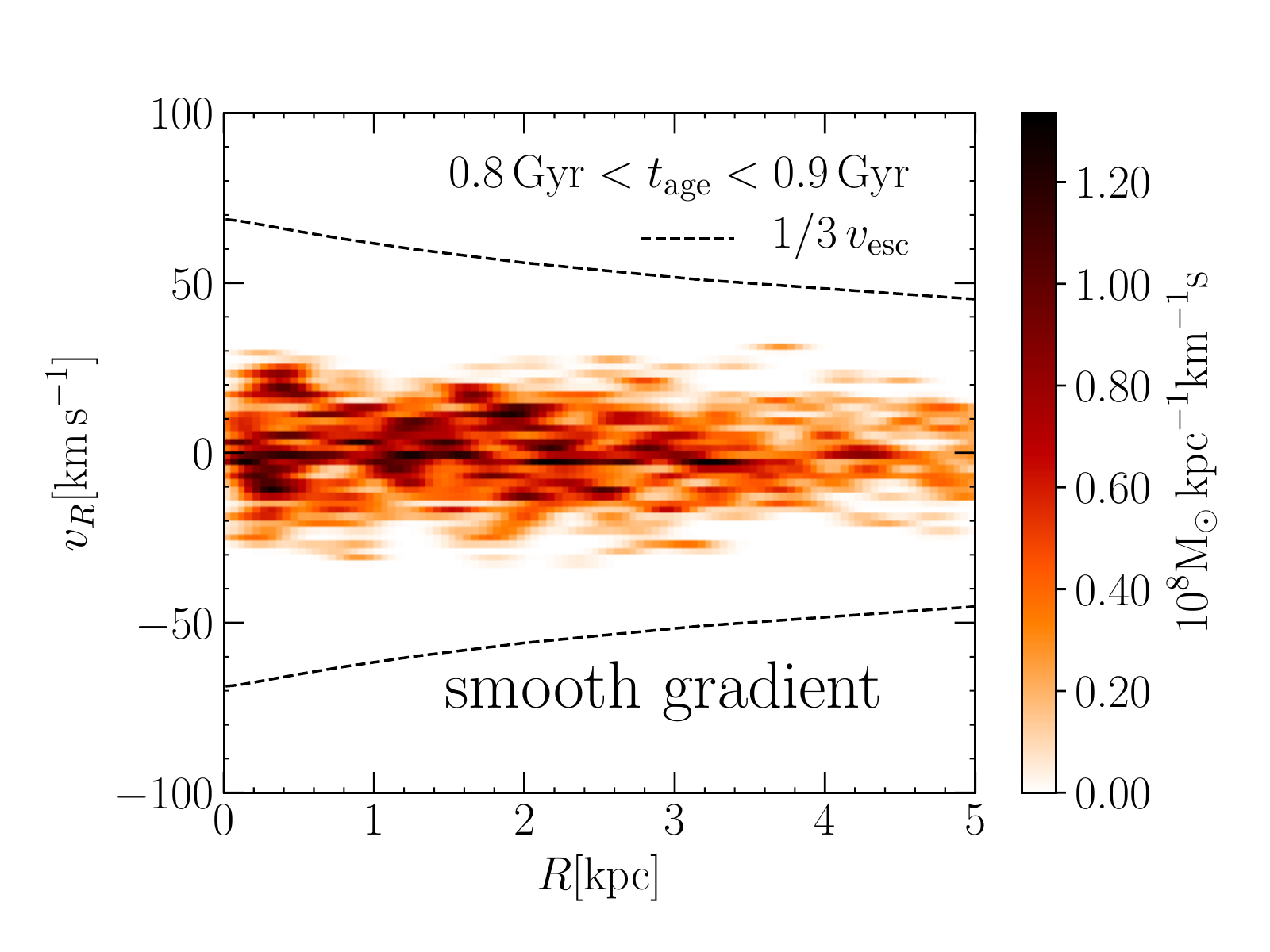}
    \includegraphics[width=8cm,trim={0.6cm 0.6cm 0.5cm 1cm},clip=true]{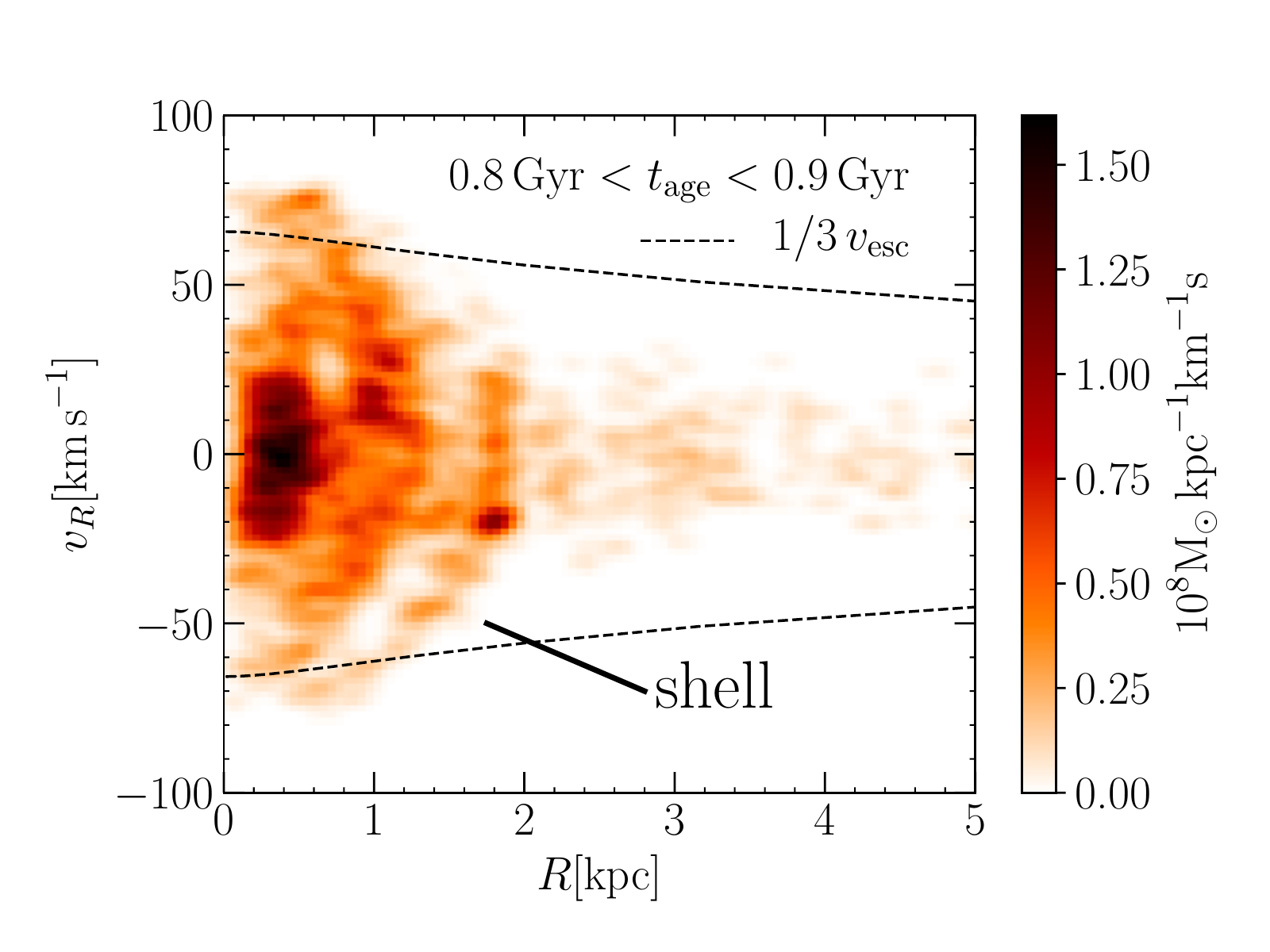}
    \caption{Average metallicity distribution (top panels) and mass-weighted distribution function (bottom panels) of star particles calculated after $3\,$Gyr of simulation time for two different simulations, projected into the radial phase space ($R-v_R$). Star particles are subject to a cut in stellar age, only 
    stars which are $0.8-0.9\,$Gyr old are shown in these plots. 
    Averaged stellar metallicity (mass-weighted distribution function) is color-coded according to the scale on the right of each panel.
    For scale, we show a third of the escape velocity as a function of radius as black dashed lines. 
    The left (right) column corresponds to the CDM simulation with smooth (bursty) star formation. 
    Smooth star formation results in a smooth metallicity gradient with the enriched/younger starts in the centre. Bursty star formation results in
    an overall weaker gradient, along with the presence of a shell of stars with high metallicity that intersects a low metallicity population at $R \sim 2$kpc. This shell appears as a distinct overdense region in the mass-weighted distribution function.}
    \label{fig:features_main}
\end{figure*}

\section{Main}
\subsection{Introduction}
One of the most persevering small-scale challenges \cite{Bullock2017} to the collisionless cold dark matter (CDM) paradigm concerns the inner density profiles of DM halos that host dwarf galaxies. Hints for constant density cores observed in some dwarf galaxies \cite{Moore1994,Kuzio2008,2019MNRAS.484.1401R,Walker2011,Oh2011,Oh2015} appear to be at odds with the ubiquitous cusps predicted by CDM $N$-body simulations \cite{1996ApJ...462..563N,2020Natur.585...39W}. To reconcile the success of the CDM paradigm at predicting the properties of the large-scale structure of the Universe with these observations on the scale of low-mass galaxies, 
a physical mechanism is required to remove the central DM cusps predicted by CDM $N$-body simulations \cite{1996ApJ...462..563N,2020Natur.585...39W}. A potential way to flatten the central density profile of halos is through strong and impulsive fluctuations in the gravitational potential caused by supernova-driven episodes of gas removal \cite{Navarro1996,Pontzen:2011ty,DiCintio2014,2019MNRAS.488.2387B,2020MNRAS.497.2393L,2020MNRAS.499.2648D,2021ApJ...921..126B}.
For SNF to be effective, supernovae must
occur in quasi-periodic cycles 
and cause strong fluctuations in the central potential on timescales that are shorter than the typical dynamical time in the galaxy, i.e., impulsively \cite{Pontzen:2011ty}. As the time between the birth of a heavy ($\gtrsim 8{\rm M_\odot}$) star and its type II supernova explosion is less than $\sim 40$ Myrs, such quasi-periodic SN cycles can only be realized in galaxies with a bursty star formation history, i.e. a star formation rate that shows order of magnitude fluctuations over less than a dynamical time (see also \cite{2019MNRAS.486.4790B}). Although there is evidence of bursty star formation in dwarf galaxies at the high mass end \cite{Kauffmann2014}, the duration of star bursts in the intermediate and low mass regime is far more uncertain \cite{Weisz2014,Emami2019}. 
In cosmological simulations, SNF is most efficient at forming cores on the scale of bright dwarfs \cite{DiCintio2014,2019MNRAS.488.2387B,2020MNRAS.497.2393L}, as long as the simulated star formation history is bursty and the gas dominates the binding energy in the inner halo before being expelled by feedback, causing a strong fluctuation in the total potential \cite{2019MNRAS.486.4790B,2020MNRAS.499.2648D}.

In modern sub-resolution models of the interstellar medium (e.g.\cite{2012MNRAS.421.3488H,2019MNRAS.489.4233M}), both the (average) burstiness of star formation and the maximal densities to which gas cools before forming stars
can be regulated via a single numerical parameter set at the resolved scales in the simulations,
the so-called SF threshold $n_{\rm th}$, which gives the minimum density that gas needs to reach before it is eligible to form stars \cite{2020MNRAS.499.2648D}. In an otherwise fixed, idealized (non-cosmological) setup, adopting larger values of $n_{\rm th}$ results in burstier SF, more substantial potential fluctuations, and thus more impulsive SNF, until eventually a threshold for core formation is reached (see \cite{2021ApJ...921..126B} for a discussion).  

An adiabatic way to form a core is through elastic scattering between DM particles. Self-interacting DM (SIDM, \cite{Spergel2000,Vogelsberger2012,Rocha2013,Tulin2018}) redistributes energy outside-in, leading to the formation of a 
$\sim1$~kpc size DM core in dwarf-size halos, provided the self-interaction cross section is $\sigma_T/m_\chi \sim 1\,{\rm cm^2g^{-1}}$ on the scales of dwarf galaxies 
\cite{Vogelsberger2012,Peter2013,Kaplinghat2016}. Such value of the cross section also evades current constraints \cite{Robertson2017,Robertson2018,Read2018}, while for cross sections smaller by about an order of magnitude, SIDM is indistinguishable from CDM \cite{Zavala2013}. Contrary to SNF, SIDM causes the formation of cores in all haloes below a certain mass and, thus, observations of dwarf galaxies with cuspy host haloes (e.g. \cite{Read2018}) are more challenging to explain in SIDM (however, see \cite{2019PhRvD.100f3007Z} for a possible explanation).

In this \textit{Letter} we use a suite of hydrodynamical simulations of an isolated dwarf galaxy to demonstrate that impulsive SNF produces distinct, shell-like kinematic signatures that appear in the phase space distribution of stars in dwarf galaxies -- and argue that the systematic absence of such features across star-forming dwarf galaxies with confirmed cores, and in particular in dwarfs with recent starbursts, would point to an adiabatic core formation mechanism, such as SIDM. 

\subsection{Simulations}
The results presented here are derived from a suite of 16 high-resolution (with a DM particle mass $m_{\rm DM}\sim 1.3\times 10^{3}{\rm M_\odot}$ and a typical baryon mass $m_{\rm b}\sim 1.4\times 10^{3}{\rm M_\odot}$, see Section \ref{supp_a}) simulations of an isolated dwarf galaxy with a total baryonic mass of $M_b = 7.2\times 10^8\,{\rm M_\odot}$ and structural properties similar to those of the Small Magellanic Cloud (see \cite{2012MNRAS.421.3488H,2019MNRAS.489.4233M,Burger2021} and Sections \ref{supp_a} and \ref{supp_b}). We use the formalism described in \cite{2003MNRAS.339..289S} to generate initial conditions of a system in approximate hydrostatic equilibrium.
We then simulate the evolution of the isolated system, using the ISM model \texttt{SMUGGLE} \cite{2019MNRAS.489.4233M} for the cosmological simulation code \texttt{AREPO} \cite{Springel:2009aa}, along with the SIDM model presented in \cite{Vogelsberger2012}, for 16 different combinations of the star formation threshold $n_{\rm th}$ and the SIDM self-interaction cross section $\sigma_T/m_\chi$. In Fig.~\ref{fig:dm_profiles}, we demonstrate that nearly identical constant density DM cores can form adiabatically in SIDM simulations with smooth star formation histories (low values of $n_{\rm th}$) and $\sigma_T/m_\chi \sim 1\,{\rm cm^2g^{-1}}$ and through impulsive SNF in CDM simulations with bursty star formation histories (high values of $n_{\rm th}$). 
While identical cores can form through SIDM and impulsive SNF, the phase space distribution of baryons is distinctly different between simulations with or without impulsive SNF, irrespective of whether DM is self-interacting. Hereafter, 
we illustrate this difference by comparing the results of
the CDM runs with $n_{\rm th} = 0.1\,{\rm cm^{-3}}$ (representative of smooth SF and thus adiabatic SNF) and $n_{\rm th} = 100\,{\rm cm^{-3}}$ (representative of bursty SF and thus impulsive SNF). 

Projections of the gas distribution 
after $3\,$Gyr of simulation time are shown in Fig.~\ref{fig:galaxyview} for both benchmark simulations. The gas distributions of the two runs look strikingly different, in particular towards the centre of the galaxies. In the simulation with bursty SF -- and thus with impulsive SNF -- 
the central gas density is lower than in the immediate surroundings due to a supernova-driven gas outflow extending out of the galactic disc, which can be clearly appreciated as a nearly spherical bubble in the edge-on projection. 
In contrast, in the simulation with smooth SF, the edge-on projection of the gas appears rather regular and the face-on projection has no distinct features.

\subsection{Stellar phase space shells}
Fig.~\ref{fig:features_main} shows (for both benchmark CDM runs) the metallicity distribution and the mass-weighted distribution function of mono-age stars which are $0.8-0.9\,$Gyr old 
projected onto the $R-v_R$ plane at the end of the simulation. 
For the simulation with smooth SF, 
we observe a steady decrease of the average metallicity of stars with increasing cylindrical radius, a natural consequence of the centrally concentrated star-forming gas. 
Statistically, more stars form in environments with higher gas densities, i.e., towards the centre of galaxies. Thus, the subsequent SNF cycles cause a metal enrichment of the ISM that is larger in the central regions.
Therefore, stars of subsequent generations (like the ones shown in Fig.~\ref{fig:features_main})
acquire a negative metallicity gradient.
Moreover, the radial velocities of stars are rather small in magnitude, $v_R\sim 25\,{\rm km\,s^{-1}}$ at most.
A different picture emerges in the centre of the galaxy with impulsive SNF. Instead of a 
monotonic stellar metallicity gradient, 
a pattern of several shells in $R-v_R$ space emerges in the metallicity distribution -- and the mass-weighted distribution function -- of stars with similar ages. The shells are comprised of star particles with high metallicities, some of which move at radial speeds
of more than $50\,{\rm km\,s^{-1}}$. These high metallicity shells intersect phase space regions inhabited by more metal-poor star particles whose radial speeds are smaller on average. Such features are transient for a given group of stars, but occur at various times in the evolution, and are not
unique to the CDM run with $n_{\rm th} = 100\,{\rm cm^{-3}}$; we find them also 
in other simulations (both in CDM and SIDM), as long as the SF histories in these simulations are bursty -- and SNF is impulsive.
For our choice of initial conditions and fixed SMUGGLE parameters, the transition from smooth to bursty SF -- and thus to impulsive SNF -- happens around $n_{\rm th} = 10\,{\rm cm^{-3}}$ (see \cite{2021ApJ...921..126B}). Notice that shells appear in all simulations in which SNF is impulsive -- regardless of whether the DM is self-interacting or not.


To quantify the difference between the final stellar distributions in the bursty SF case and in the smooth SF case shown in Fig.~\ref{fig:features_main}, we estimate the likelihood of randomly finding, in the smooth SF case, an overdensity similar to that associated with the clear 
shell in the bursty SF simulation. 
We take the normalized, cumulative stellar mass distribution of the smooth SF simulation as a target distribution for random sampling and construct $10^7$ re-sampled distributions, each time drawing as many radii as there are stars in the original distribution. For each re-sampled distribution, we then search a pre-defined ``signal range'' for the largest spherical overdensity that arises as a result of Poisson sampling (see Section \ref{supp_d}). We also calculate the overdensity at the position of the shell in the bursty SF simulation. From these values, we create a distribution of global (signal region) and local (shell area) overdensities. Comparison against the measured shell overdensity in the bursty SF simulation reveals that the shell has a global (local) significance of more than $5\,\sigma$ ($3.3\,\sigma$) compared to the smooth SF case.
These are conservative estimates for the significance of the shell-like feature since they are based on the stellar density distribution only and do not take into account information on the metallicity of stars. 
Finding an overdensity of such amplitude, combined with the observation that the overdensity consists mainly of high-metallicity stars, would be a smoking gun signature of an impulsive SNF cycle following an episode of bursty SF. 

To determine how the shell-like features appear in the line-of-sight phase space of galaxies that are observed edge-on, we projected the distributions shown in Fig.~\ref{fig:features_main} into $|x|-v_y$ space (using the coordinate system defined in Fig.~\ref{fig:galaxyview}; see Supplemental Fig.~\ref{fig:features_projected} and Section \ref{supp_c} for details on how this figure was created).  In the featureless smooth SF 
case, the emerging distribution of stars 
tracks the rotation curve of the galaxy, with a monotonic decrease of (average) metallicity with distance. In the bursty SF case, 
we observe two isolated overdense clusters consisting mainly of high metallicity stars, at a distance to the galaxy's centre of $\sim 2\,{\rm kpc}$, similar to the radius at which the phase space shell appears in Fig.~\ref{fig:features_main}. We explicitly confirmed that those clusters consist of the same stars as the phase space shell. We emphasize again that similar differences between the phase space distributions of stellar particles in galaxies with or without impulsive SNF emerge when SIDM simulations are considered.

\subsection{How shells are created}\label{main_b}
The shell-like features shown in Fig.~\ref{fig:features_main} 
arise in the aftermath of starburst events -- which are closely followed by impulsive episodes of SNF. Reference \cite{2016ApJ...820..131E} showed that young stars born in a turbulent ISM inherit the orbits of the star-forming gas and can be born with significant radial motion. The orbits of these stars are then further heated by subsequent feedback episodes, leading to sustained radial migration. The shells presented here form from such groups of stars, which are born during starburst events in a turbulent ISM. Such groups of stars constitute {\it orbital families}\cite{2019MNRAS.485.1008B,2021ApJ...921..126B} -- sets of orbits defined by similar integrals of motion (see also Section \ref{sec:shell_formation}). Moreover, they are born with similar metallicities and -- as outlined above -- with some initial amount of radial motion.

Instead of causing a coherent net expansion, subsequent impulsive fluctuations in the gravitational potential discontinuously change the (gravitational) energy of a star (particle) 
by an amount that depends on its orbital phase \cite{Pontzen:2011ty,2019MNRAS.485.1008B}. As a consequence, they can split an initially phase mixed {\it orbital family}, i.e., create a distribution that is unmixed \cite{2021ApJ...921..126B}. The phase space shells we observe are therefore signatures of the early stages of phase mixing \cite{2008gady.book.....B,2011EPJP..126...55D}. To compare this to the smooth case, we note that in dynamical systems in which orbits are regular and stars act as dynamical tracers of the gravitational potential, the average metallicity of stars can only depend on their actions \cite{2021ApJ...910...17P}. We can therefore generally assume {\it orbital families} to be well approximated by groups of stars with similar ages and metallicities. Across our simulation suite, we find that in simulations with impulsive SNF (following bursty SF), 
the energy distribution of {\it orbital families} is wider than in simulations with smooth SF (see Supplemental Fig.~\ref{fig:hists} and related discussion in Section \ref{supp_c}), a direct result of the periodic, SNF-driven heating of stellar orbits (see \cite{2019MNRAS.485.1008B}). 
The shell-like signatures of early-stage phase mixing observed here are thus a direct consequence of impulsive SNF.


\subsection{Discussion and outlook}
Finding stellar shells similar to the ones presented here in nearby dwarfs would imply a prior episode of impulsive SNF without necessarily establishing SNF as the dominant core formation mechanism.
In cosmological halos, we expect that diffusion caused by the halo's triaxial shape will erase shells within $\sim 1$ dynamical time \cite{2019MNRAS.485.1008B}, implying that galaxies without recent bursty star formation (e.g. quenched galaxies) are fairly bad targets to look for phase space shells. 
Nevertheless, it is instructive to evaluate the potential of detecting such signatures of bursty SF in the Milky Way satellites, in particular  
Fornax \cite{Walker2009,deBoer2012,Rusakov2021}, since it has been claimed to have a core \cite{Walker2011} (albeit this remains controversial, see \cite{Genina2018}) and information on line-of-sight kinematics, metallicity \cite{Walker2009}, and age \cite{deBoer2012} is available for a sub-sample of 
its member stars. Unfortunately, we find that the ages of individual stars carry uncertainties which are too large ($\sim$ 1 Gyr)
to conclusively identify {\it orbital families}. Ideal future targets to look for impulsive SNF signatures are star-forming field dwarfs in the vicinity of the Local Group (see \cite{2019MNRAS.484.1401R}). At the current time, the number of resolved stars with known ages and metallicities in these galaxies is too small. Within the next decade, however, the Roman Space Telescope will provide precise photometric data of individual stars in dwarf galaxies within the Local Volume \cite{Khan2018}. 
Combined with spectroscopic data from the ground, this will enable the precision needed to determine the ages, metallicities, and kinematics of a sufficient number of stars to conclusively establish whether the characteristic shell-like signatures of impulsive SNF presented here -- or rather their projections into the space of line-of-sight velocity vs projected radius (see Supplemental Fig.~\ref{fig:features_projected}) -- are ubiquitously present or systematically absent. Further in the future, new generations of extremely large telescopes (ELTs) may even provide sufficiently precise data on the 3d motions of stars in nearby dwarfs to allow for a search of shells directly in $R-v_R$ space \cite{2019BAAS...51c.153S}.

The significance of a (potential) non-detection of such shells in dwarf galaxies with a core also depends on how robust our results are to changes in the initial setup or the stellar evolution model. Apart from the host halo's triaxiality, two effects that we do not explicitly test for may be significant. First, SF histories in real dwarf galaxies may be bursty, but starbursts may occur away from the galaxy's center. However, SNF needs to impulsively change the central potential to be a feasible core formation mechanism (see \cite{2021ApJ...921..126B,Pontzen:2011ty}). The non-detection of kinematic signatures would then require starbursts to occur mainly in the center of dwarfs, but exclusively off-center at late times; a possible but unlikely scenario (see \cite{2016ApJ...820..131E}). Second, stellar clusters (i.e. {\it orbital families}) born in starburst events need to contain a sufficient number of stars to allow us to identify shells formed from them. Based on our re-sampling routine (see Section \ref{supp_d}), we estimate that a shell needs to contain a few hundred stars to be significant at a $2\,\sigma$ level (notice that this implies that future experiments need to provide precise age and metallicity information for $\mathcal{O}(10^4)$ stars if we assume a constant average SF rate). In a detailed high-resolution study of the impact of different kinds of stellar feedback, \cite{Smith2021} found that clusters of such size formed in all simulations in which SNF was included. At this time, we are unaware of any study in which no clusters of at least $\sim 100$ stars form while SNF is modeled self-consistently and found to be a feasible core formation mechanism.
We thus infer that the conclusive, systematic absence of signatures of impulsive SNF across all isolated, star-forming field dwarfs with confirmed cores would give strong support to alternative DM models, such as SIDM.

\setcounter{figure}{0}
\renewcommand{\figurename}{Supplemental Fig.}

\begin{figure}
    \centering
    \includegraphics[width=\linewidth]{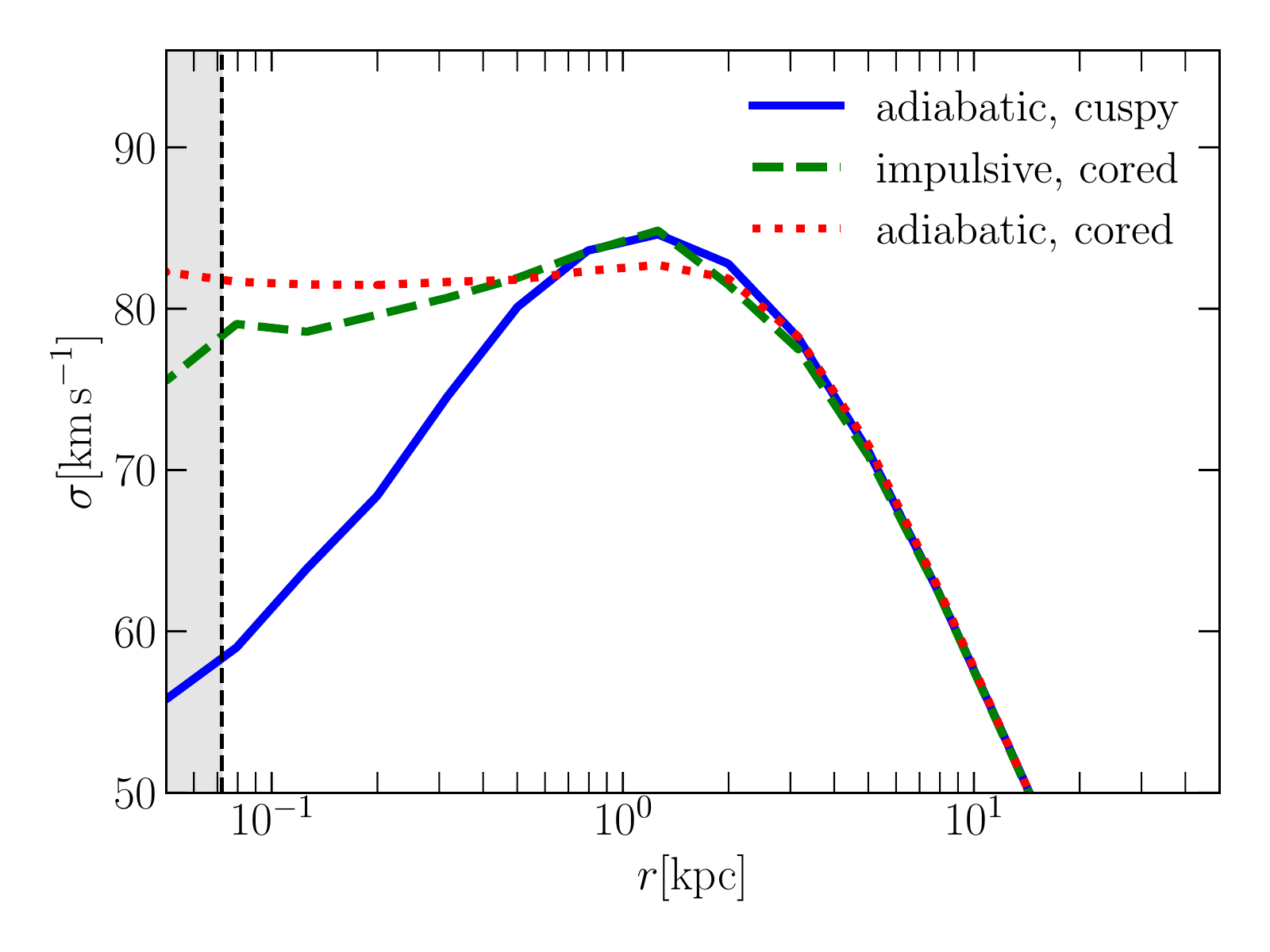}
    \caption{Adiabatic and impulsive cusp-core transformation: Spherically averaged velocity dispersion profiles of three different simulated dwarf-size halos after $3\,$Gyr of simulation time. The solid blue line corresponds to the CDM simulation with smooth star formation ($n_{\rm th} = 0.1\,{\rm cm^{-3}}$), whereas the dashed green line denotes results of the CDM simulation with bursty star formation ($n_{\rm th} = 100\,{\rm cm^{-3}}$). 
    The red dotted line shows the SIDM simulation with smooth star formation ($n_{\rm th} = 0.1\,{\rm cm^{-3}}$) 
    and a self-interaction cross section $\sigma_T/m_\chi = 1\,{\rm cm^2g^{-1}}$. }
    \label{fig:dm_profiles_supp}
\end{figure}

\begin{figure}
    \centering
    \includegraphics[width=0.48\linewidth]{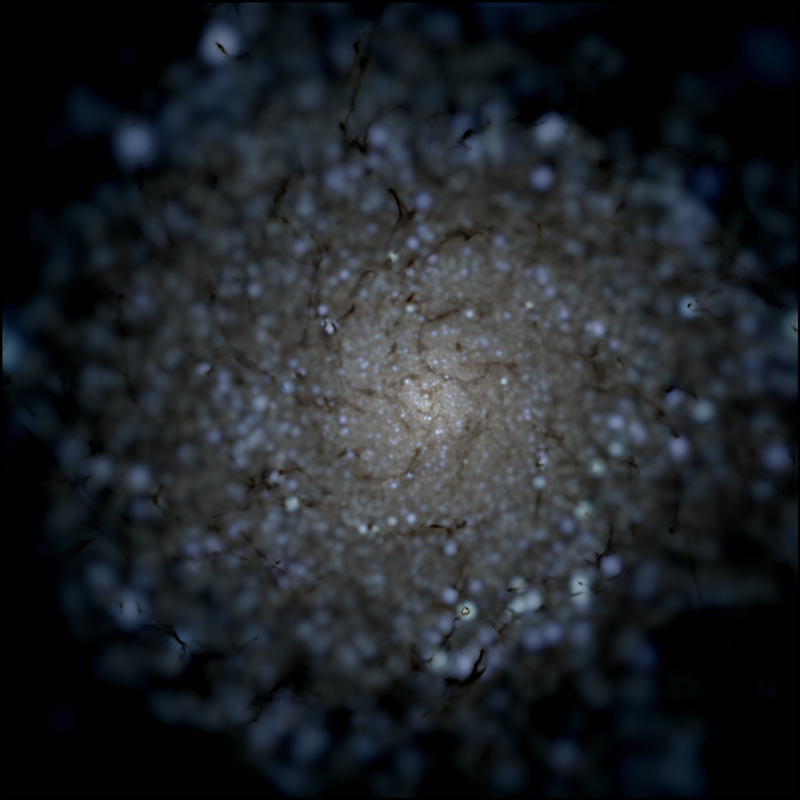}
    \includegraphics[width=0.48\linewidth]{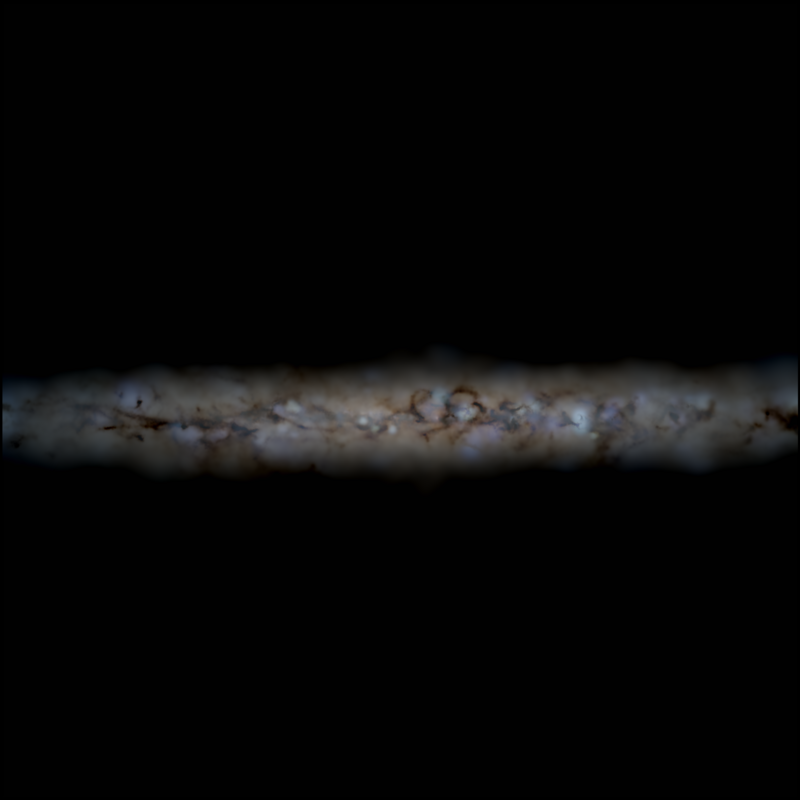}\\
    \includegraphics[width=0.48\linewidth]{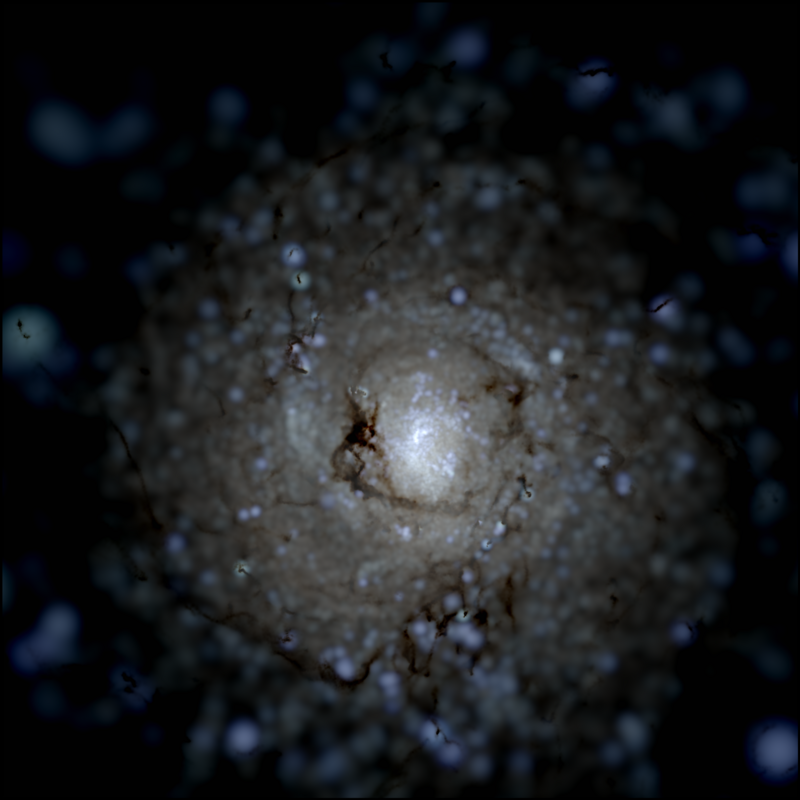}
    \includegraphics[width=0.48\linewidth]{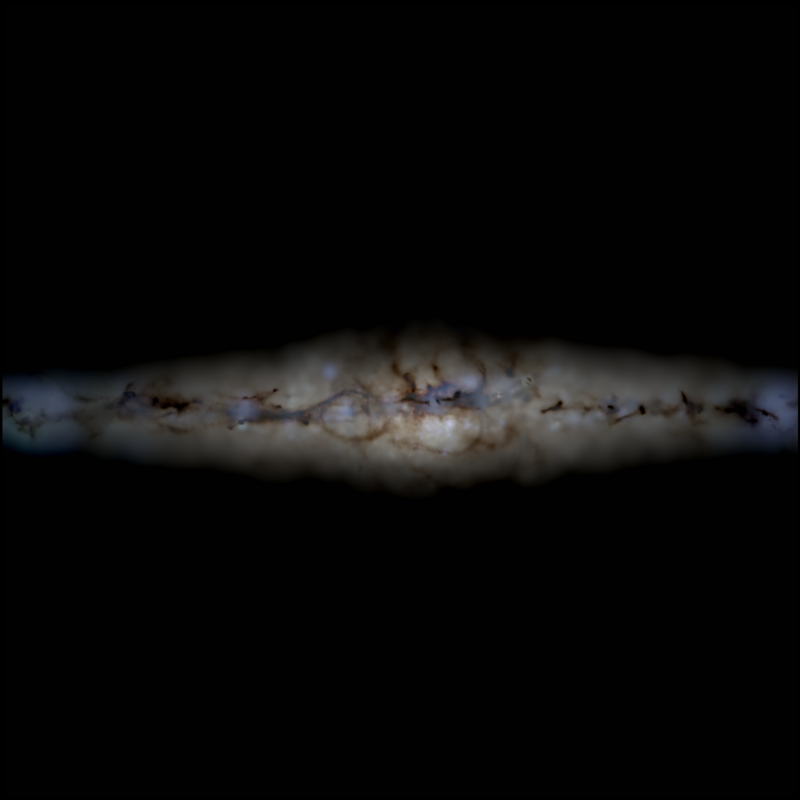}
    
    \caption{Resulting galaxies after $3\,$Gyr of simulation time for two simulations with different star formation thresholds. 
    In the top panels, we show face-on (left column) and edge-on (right column) projections of the stellar light
    for the CDM simulation with smooth (top row) and bursty (bottom row) star formation. The side-length of the field of view is $20\,$kpc in each panel.} 
    \label{fig:galaxyview_supp}
\end{figure}


\begin{figure*}
    \centering
    \includegraphics[width=8cm,trim={0.6cm 0.6cm 0.5cm 0.5cm},clip=true]{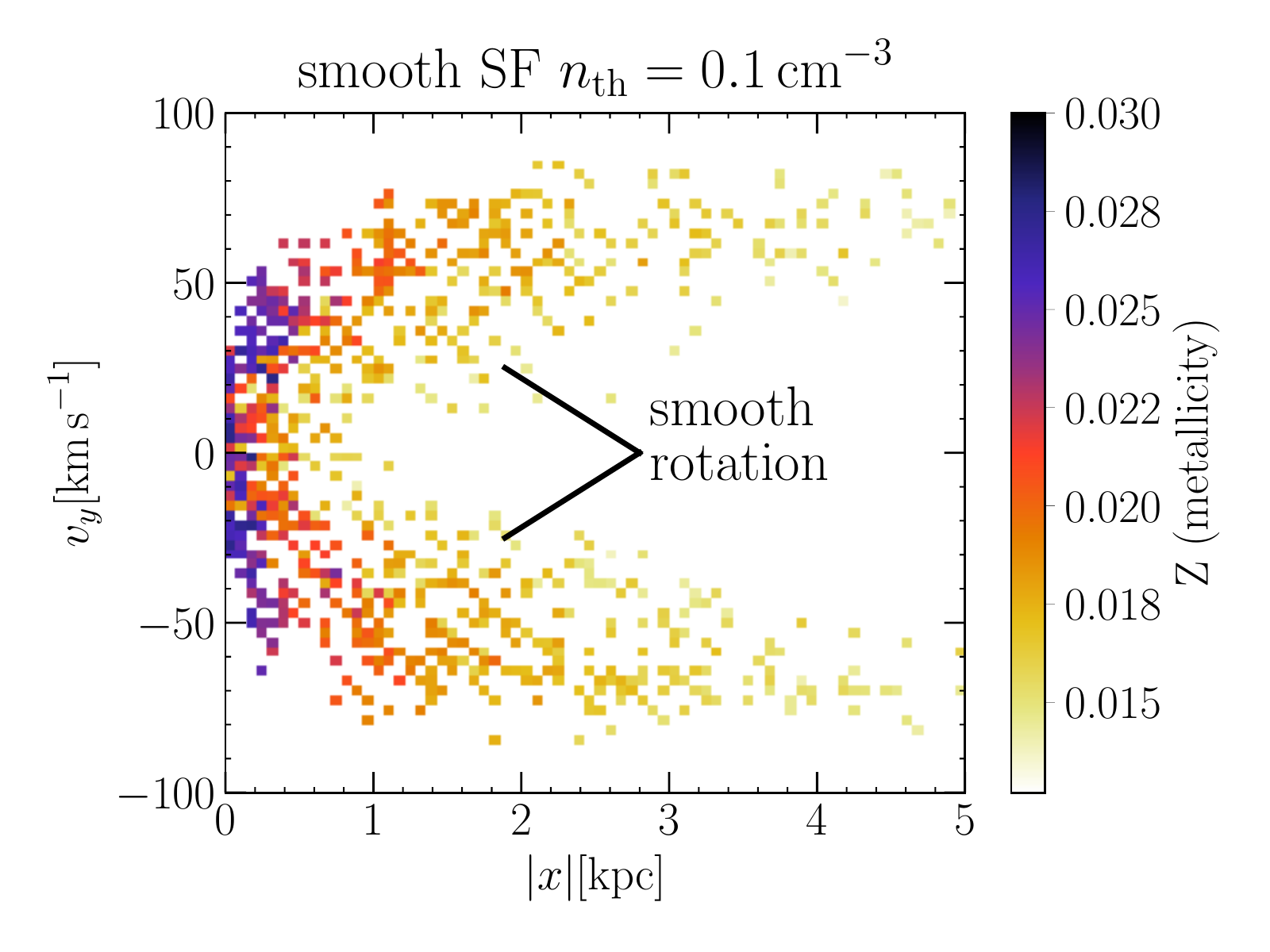}
    \includegraphics[width=8cm,trim={0.6cm 0.6cm 0.5cm 0.5cm},clip=true]{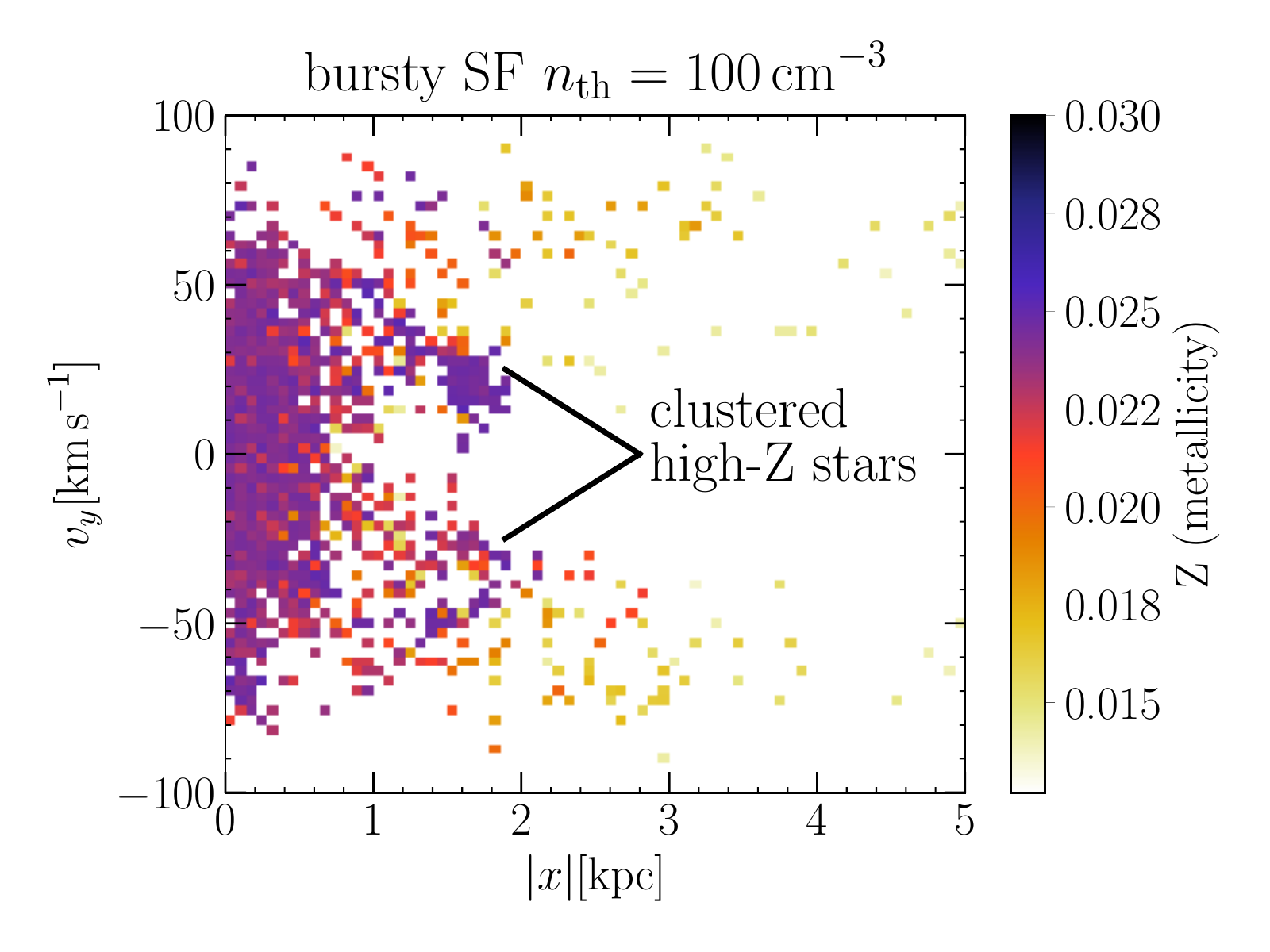}\\
    \includegraphics[width=8cm,trim={0.6cm 0.6cm 0.5cm 1cm},clip=true]{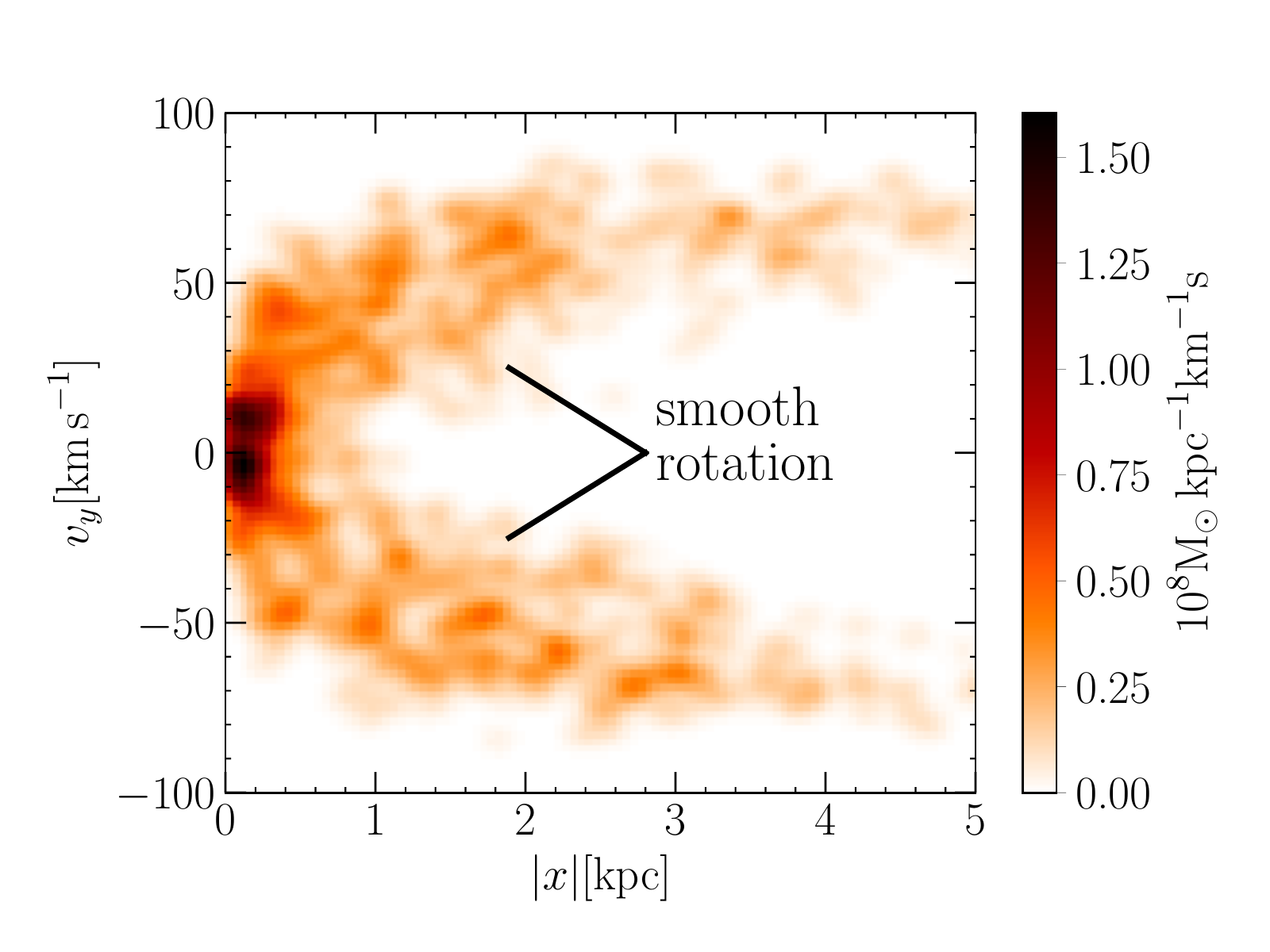}
    \includegraphics[width=8cm,trim={0.6cm 0.6cm 0.5cm 1cm},clip=true]{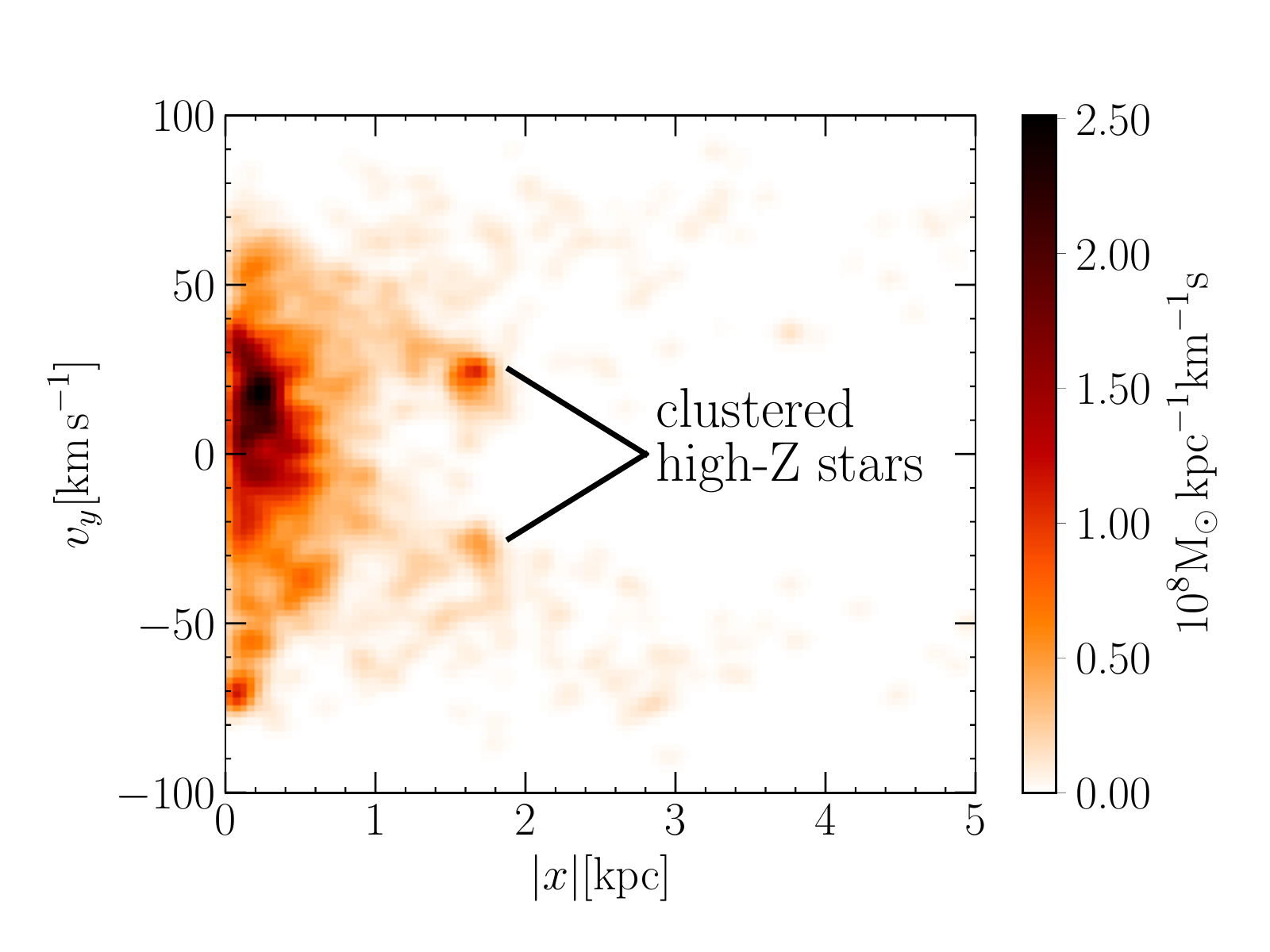}
    \caption{Average metallicity distribution (top panels) and mass-weighted distribution function (bottom panels) of star particles calculated after $3\,$Gyr of simulation time for two different simulations, projected into $|x|-v_y$ space. The same as Fig.~\ref{fig:features_main} but along the disk plane: in the $x$ direction spatially, and in the $y$ direction in velocity space. This gives an idea of the expected stellar kinematics projected into the line-of-sight phase space. Notice how in the left panels (smooth star formation), a smooth, but spread/scattered, rotation curve can be observed, while in the right panels, features associated to the impulsive case (bursty star formation), are apparent, albeit their structure is harder to discern (relative to the radial phase space shown in Fig.~\ref{fig:features_main}, where these features are clear). In particular, two distinct overdense clusters of high metallicity stars appear at $|x| \sim 2\,{\rm kpc}$ in the impulsive (bursty star formation) case.
     }
    \label{fig:features_projected}
\end{figure*}

\begin{figure*}
    \centering
    \includegraphics[width=8cm, trim={0.5cm 0.5cm 0.5cm 0.5cm},clip=true]{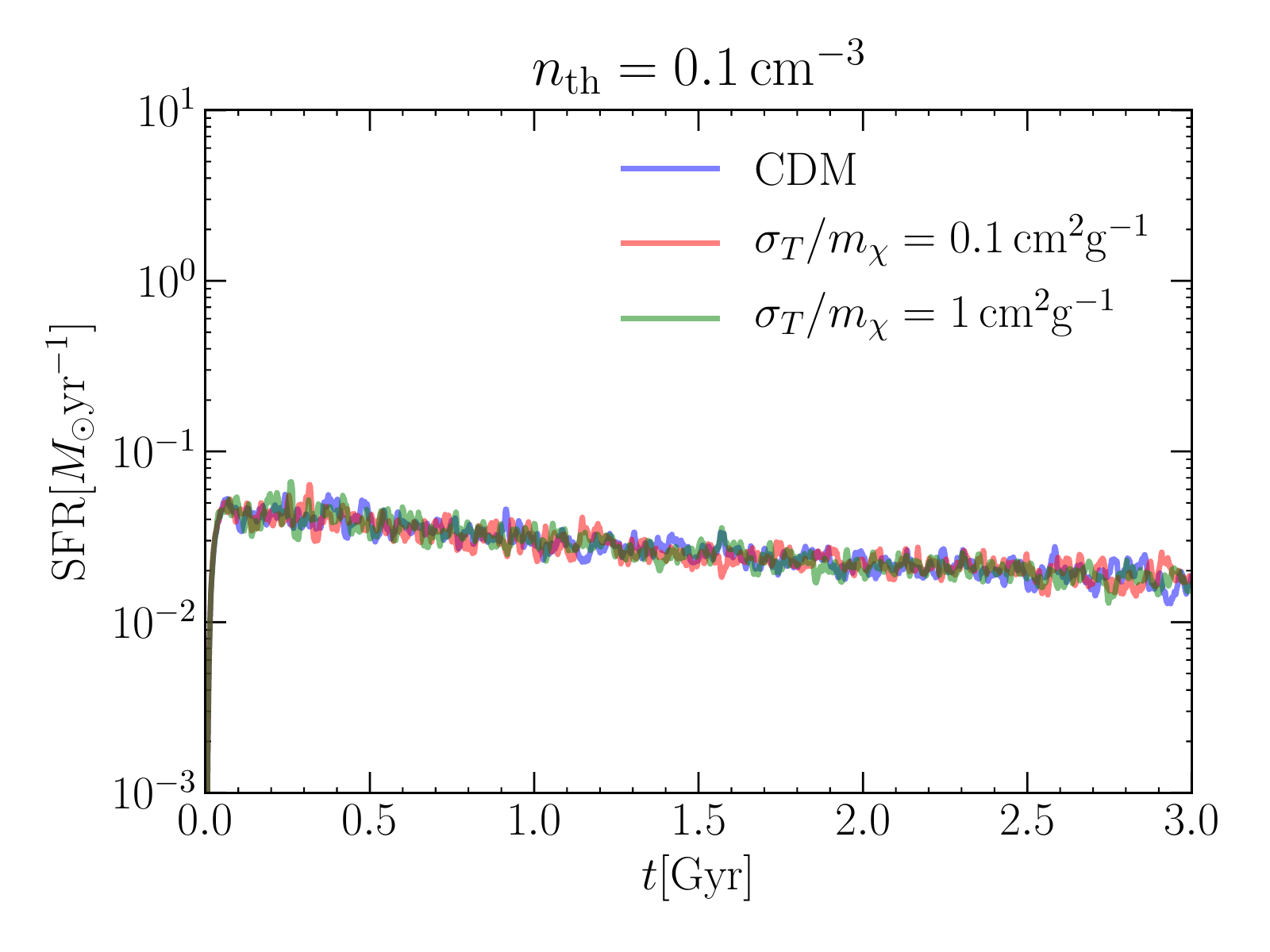}
    \includegraphics[width=8cm, trim={0.5cm 0.5cm 0.5cm 0.5cm},clip=true]{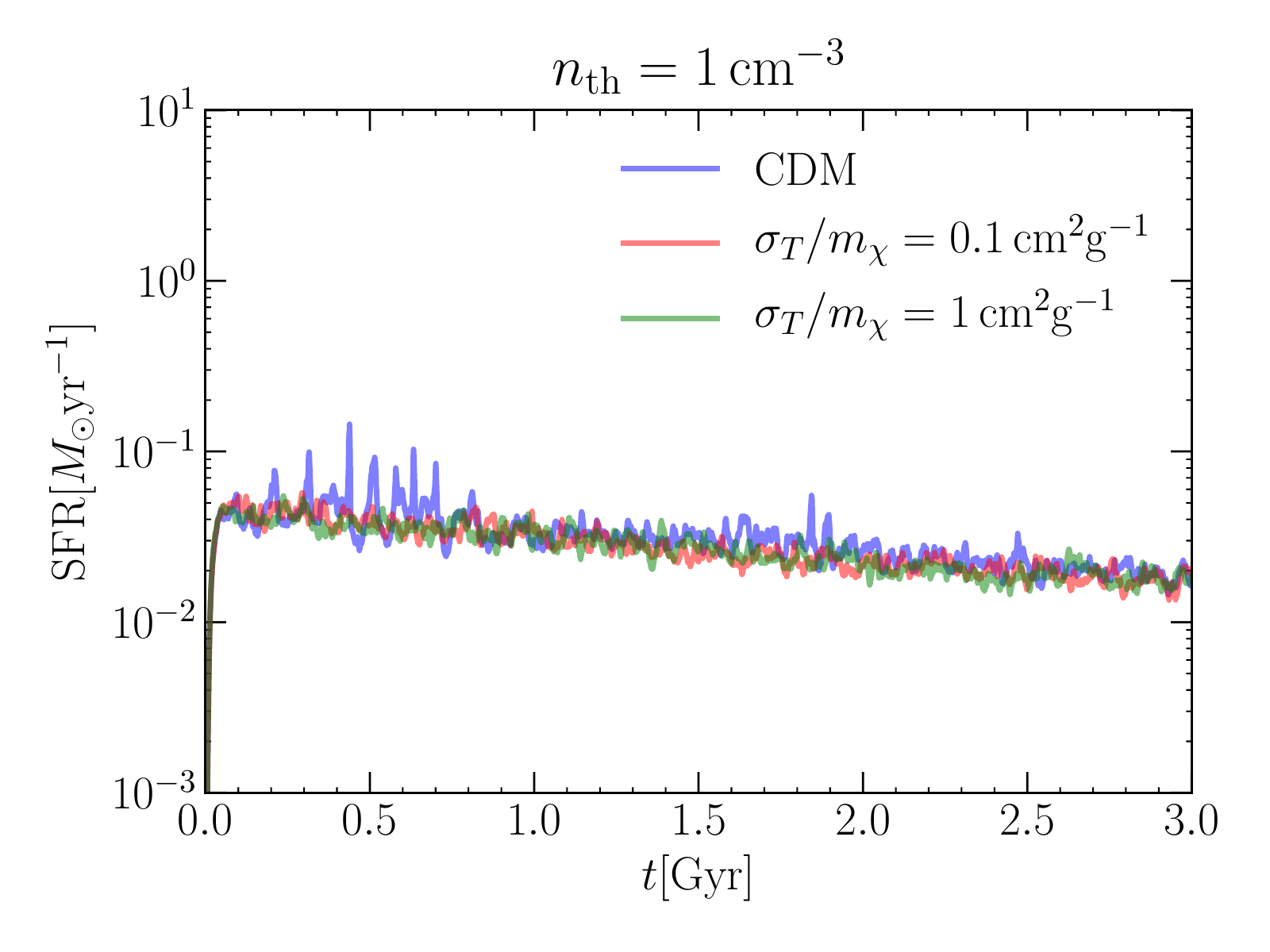}\\
    \includegraphics[width=8cm, trim={0.5cm 0.5cm 0.5cm 0.5cm},clip=true]{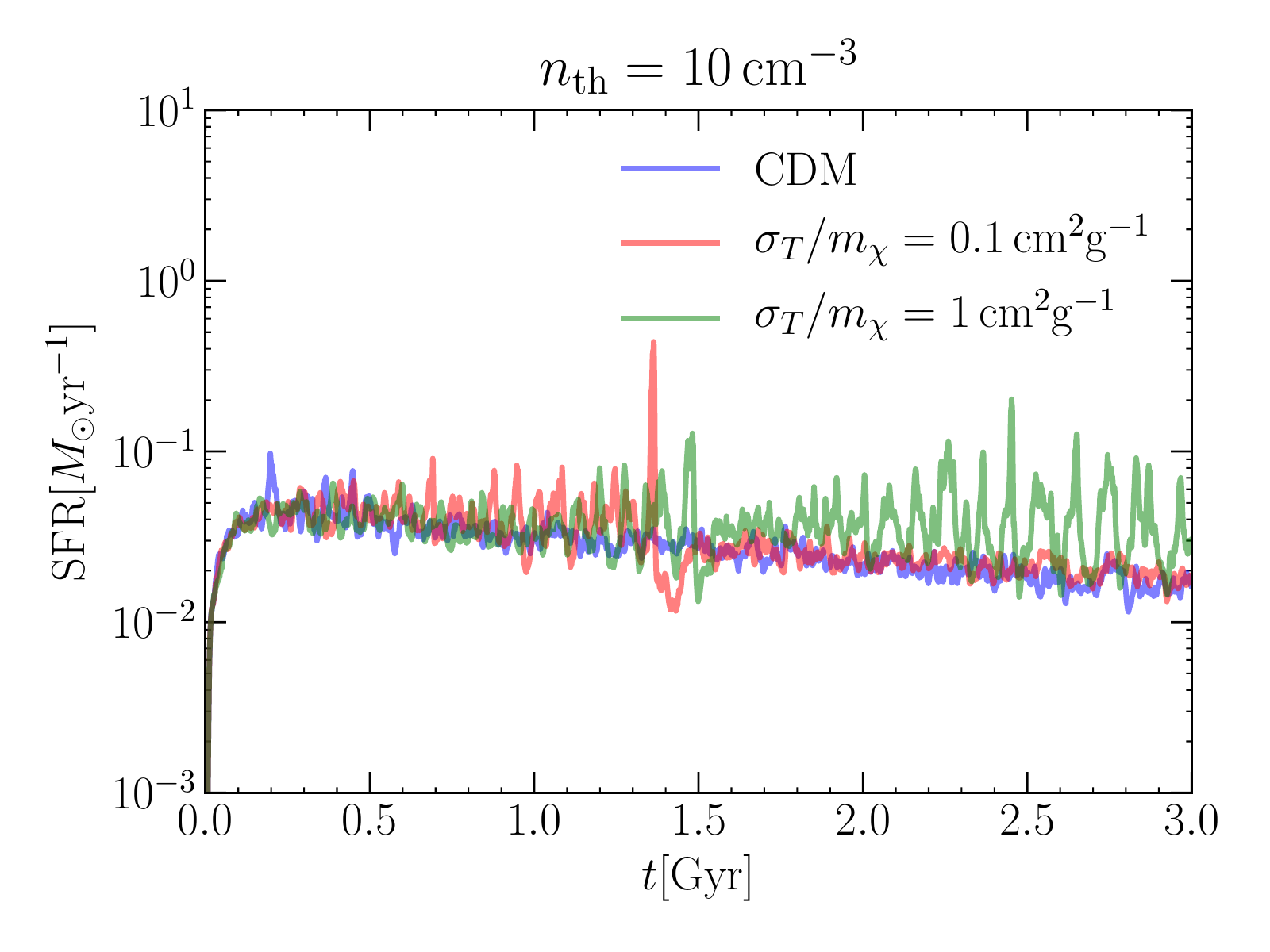}
    \includegraphics[width=8cm, trim={0.5cm 0.5cm 0.5cm 0.5cm},clip=true]{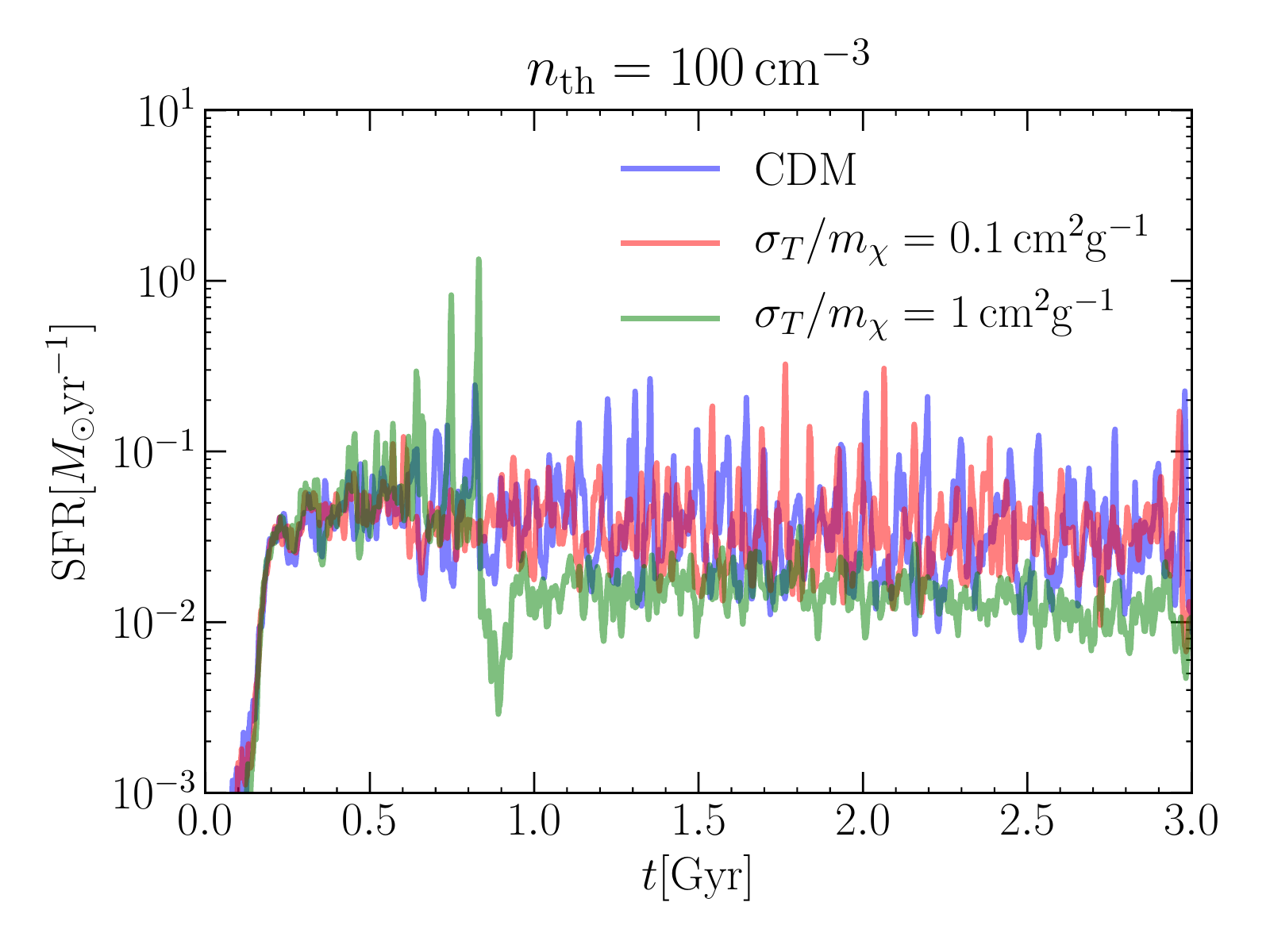}
    \caption{SF histories for 12 different simulations within CDM and SIDM. Different panels correspond to simulations with different SF thresholds, $n_{\rm th}$, according to the legend on the top of each panel. 
    Within each panel, differently coloured lines correspond to simulations with different DM self-interaction cross sections as indicated in the figure legends. As the SF threshold is increased, SF bursts become more prominent and episodic. After particularly intense bursts, the SF history can diverge across simulations due to the stochastic nature of the SF implementation. The benchmark runs of this {\it Letter} are the CDM run with $n_{\rm th} = 0.1\,{\rm cm^{-3}}$ (blue line upper left panel) and the CDM run with $n_{\rm th} = 100\,{\rm cm^{-3}}$ (blue line lower right panel). The SIDM run used in Fig.~\ref{fig:dm_profiles} and Supplemental Fig.~\ref{fig:dm_profiles_supp} has $n_{\rm th} = 0.1\,{\rm cm^{-1}}$ and $\sigma_T/m_\chi = 1\,{\rm cm^2g^{-1}}$ (green line upper left panel).}
    \label{fig:sfr}
\end{figure*}

\begin{figure}
    \centering
    \includegraphics[width=\linewidth,trim={0.5cm 0.5cm 0.5cm 0.5cm},clip=true]{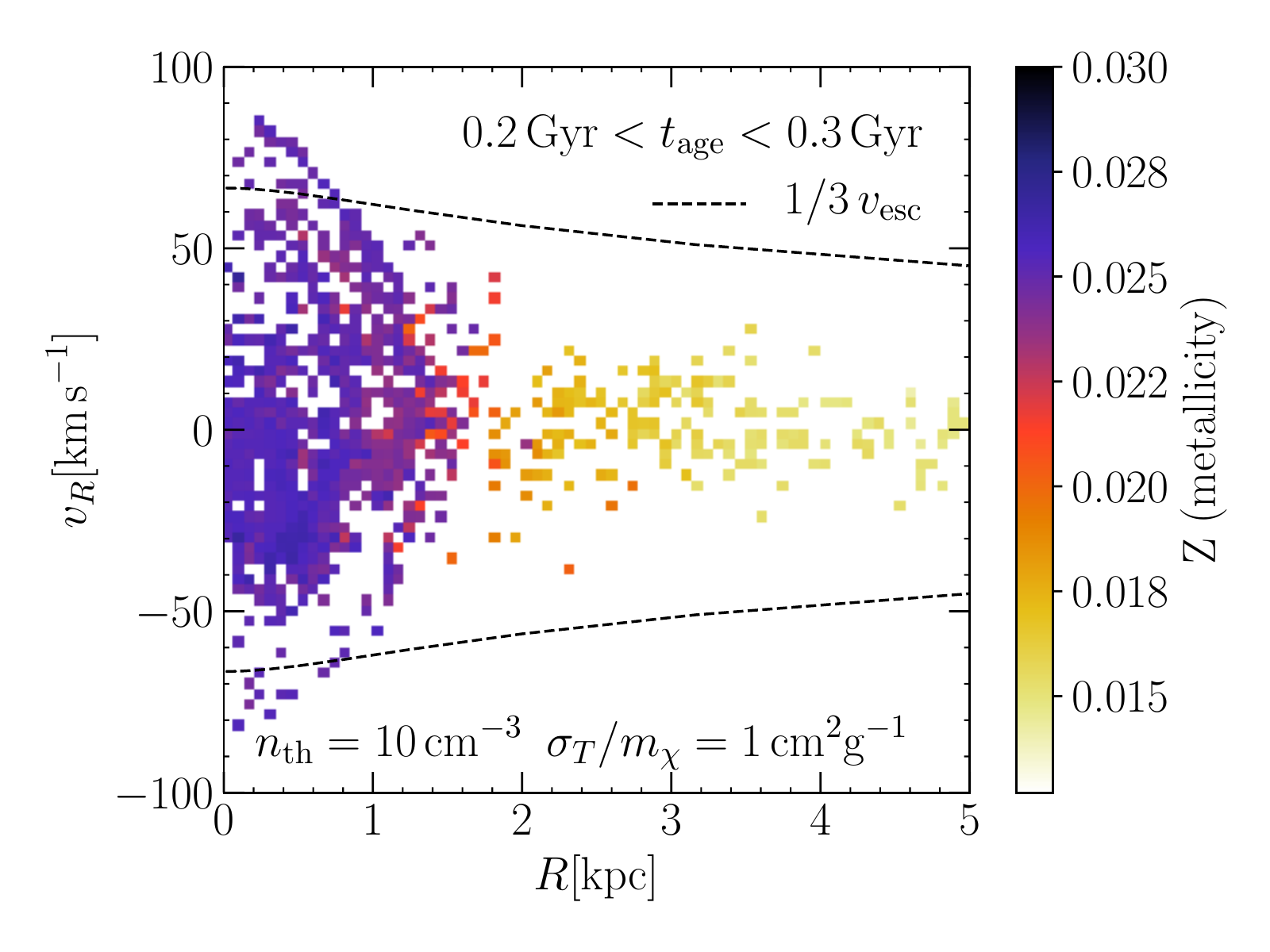}\\
    \includegraphics[width=\linewidth,trim={0.5cm 0.5cm 0.5cm 0.5cm},clip=true]{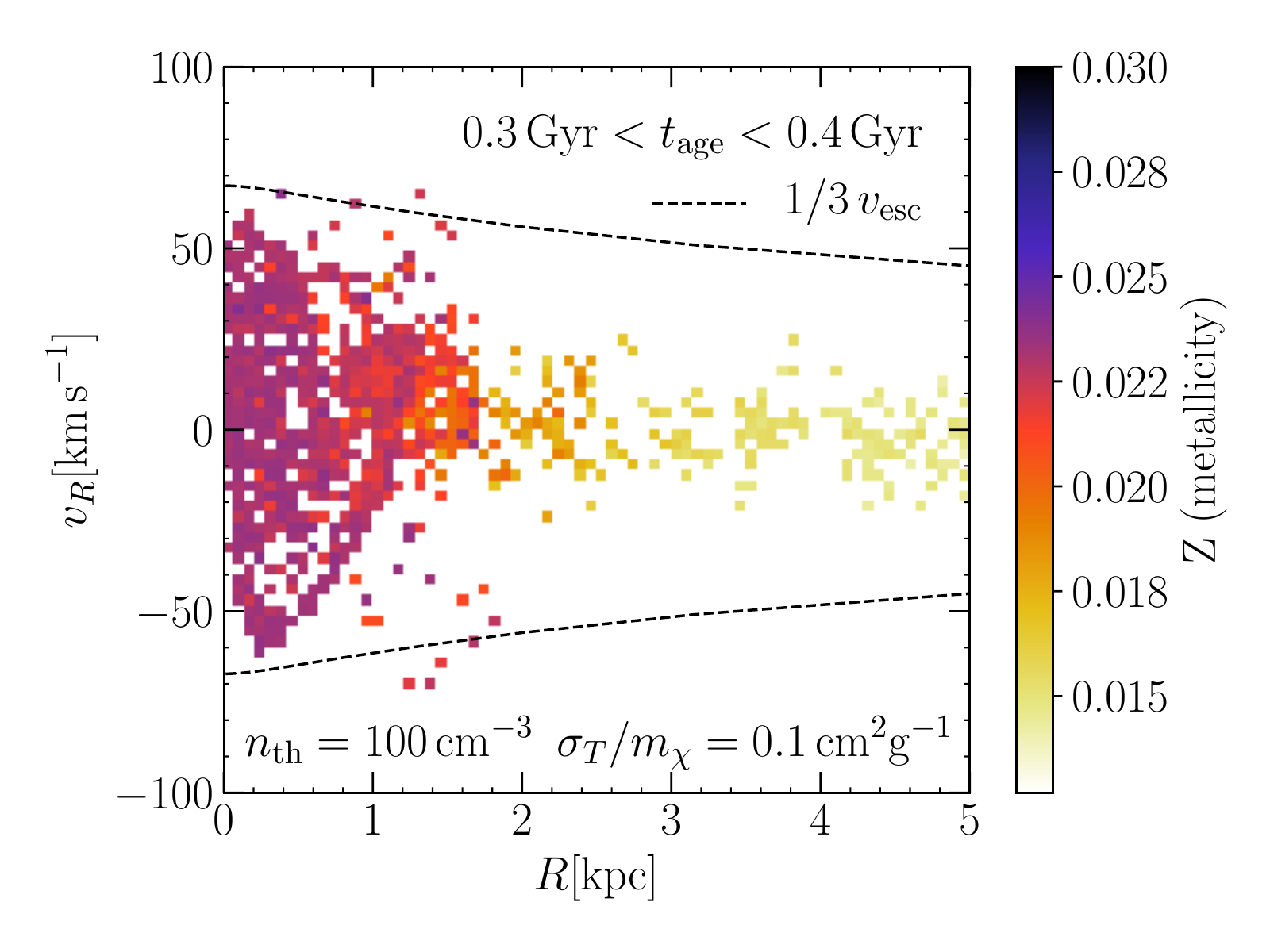}
    \caption{
    Radial phase space projections of the metallicity distribution of star particles within a given age range formed in two different simulations, calculated after different simulation times. The top panel corresponds to the simulation with ``mildly'' bursty SF ($n_{\rm th} = 10\,{\rm cm^{-3}}$) and $\sigma_T/m_\chi = 1\,{\rm cm^2g^{-1}}$ and only shows 
    stars which are $0.2-0.3\,$Gyr old, calculated from a $2.4\,$Gyr snapshot. The bottom panel corresponds to the simulation with bursty SF ($n_{\rm th} = 100\,{\rm cm^{-3}}$) and $\sigma_T/m_\chi = 0.1\,{\rm cm^2g^{-1}}$ and only shows 
    stars which are $0.3-0.4\,$Gyr old, calculated from a $2.4\,$Gyr snapshot. Stellar metallicities are averaged in $70\times 70$ bins in $R-v_R$ space and the displayed average metallicity per bin is colour-coded as indicated to the right of each panel. Shell-like features such as the ones presented here and in Fig.~\ref{fig:features_main} are found at different times across all simulations with bursty SF.}
    \label{fig:features_app}
\end{figure}

\begin{figure*}
    \centering
    \includegraphics[width=0.49\linewidth,trim={0.5cm 0.5cm 0.5cm 0.5cm},clip=true]{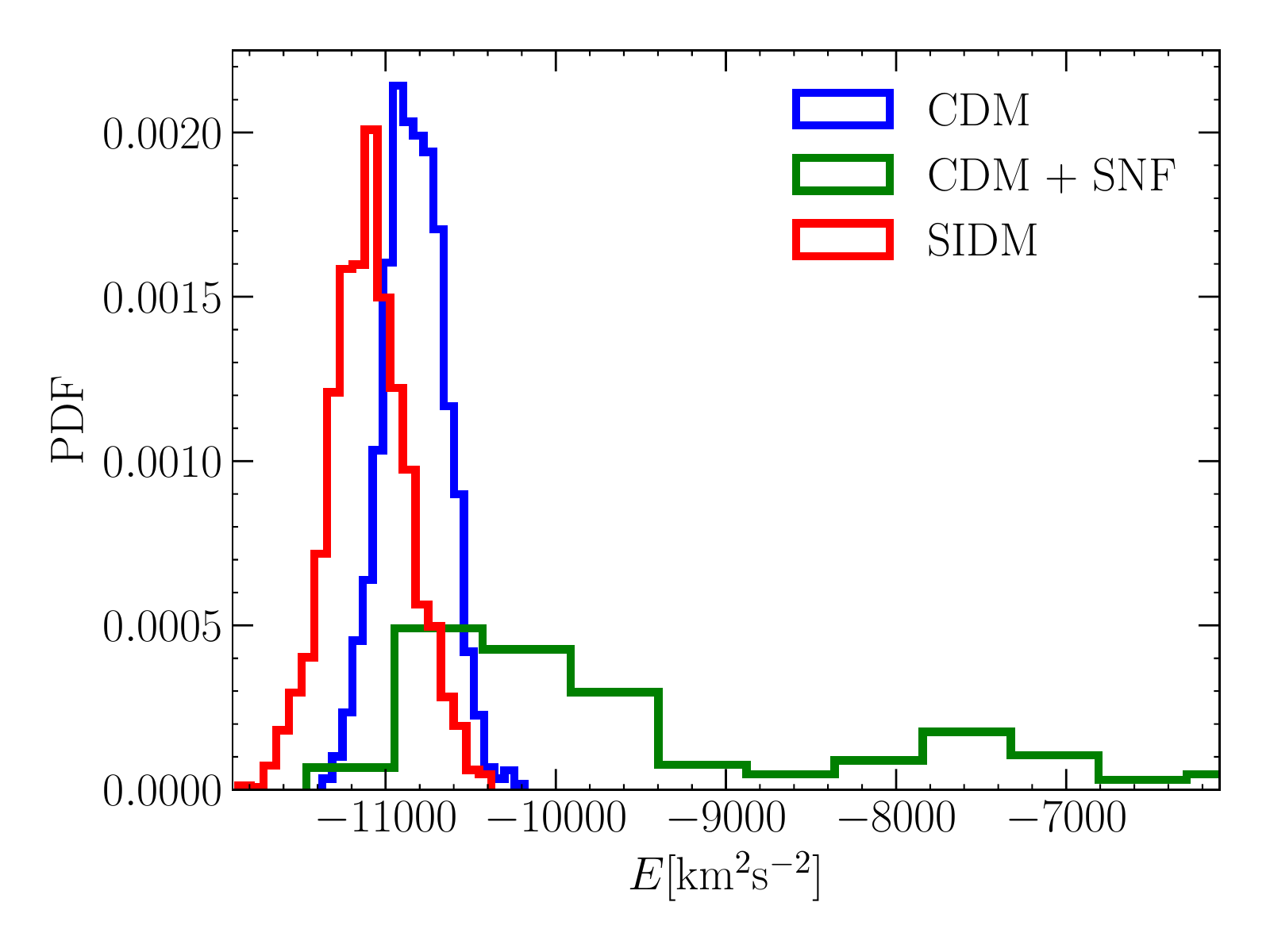}\\
    \includegraphics[width=0.49\linewidth,trim={0.5cm 0.5cm 0.5cm 0.5cm},clip=true]{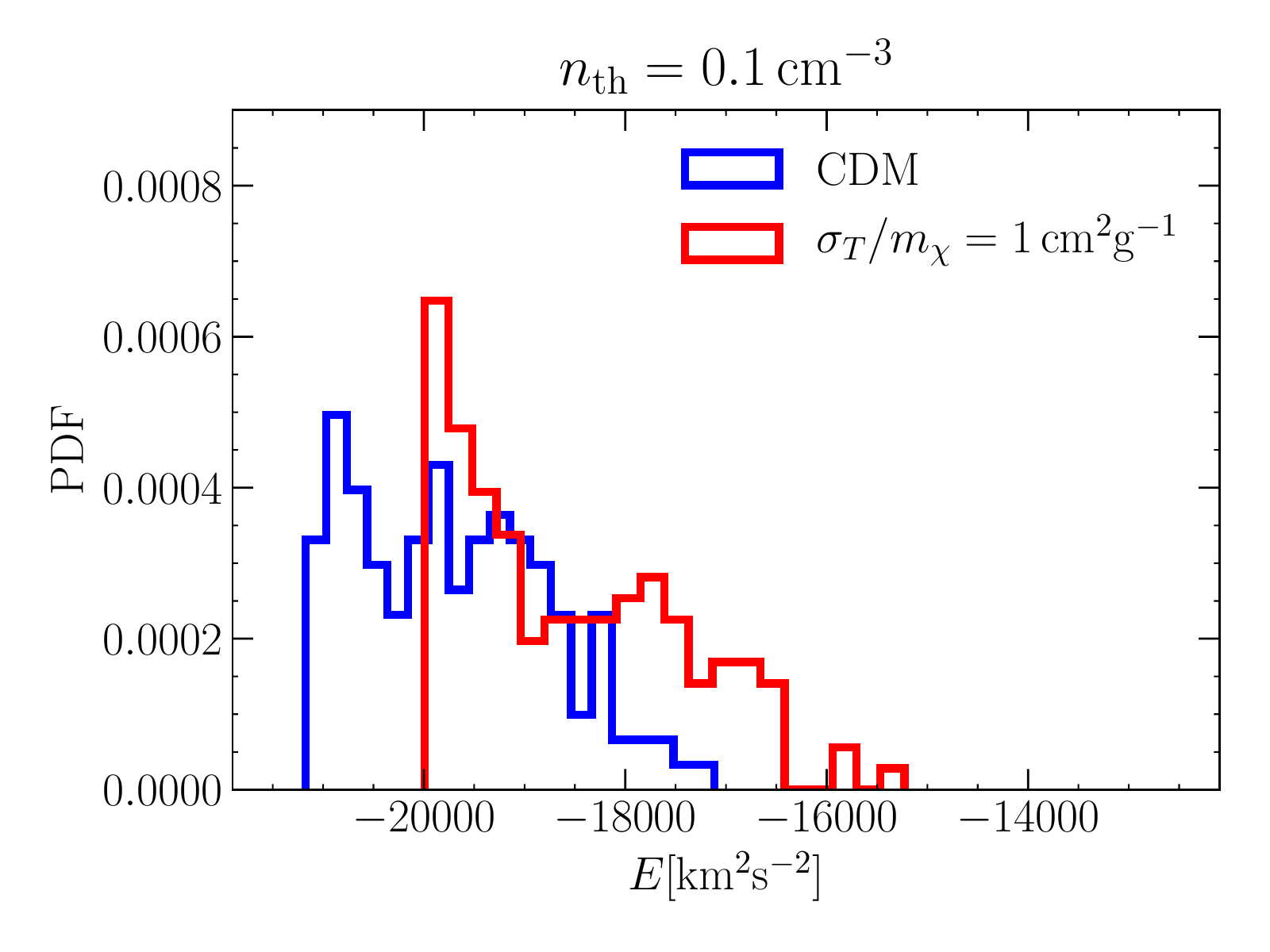}
    \includegraphics[width=0.49\linewidth,trim={0.5cm 0.5cm 0.5cm 0.5cm},clip=true]{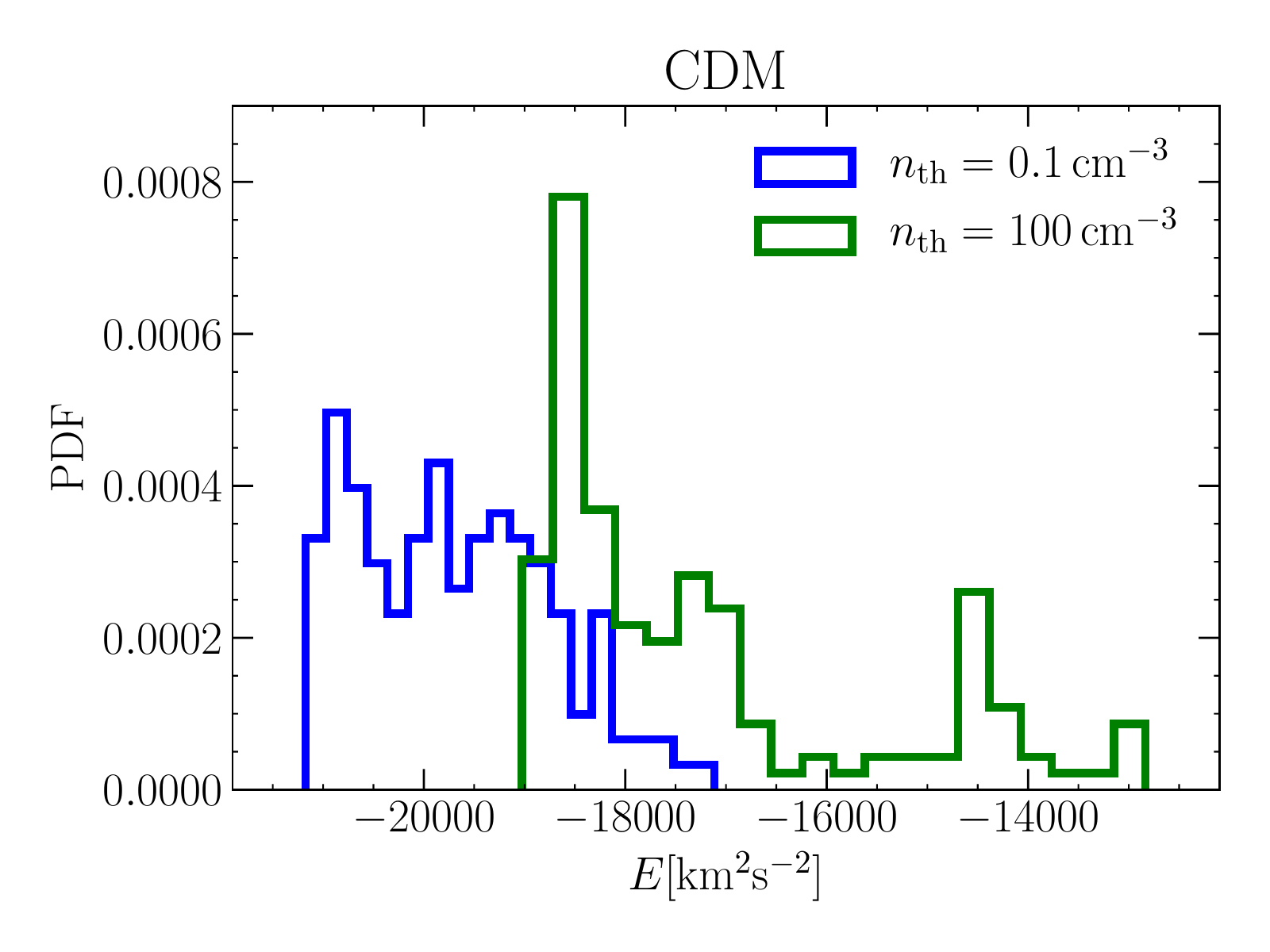}
    \caption{ Comparison of energy distributions of tracer (stellar) particles from different simulations after 3 Gyr of simulation time. The top panel shows the energy distribution of the {\it orbital families} of tracer particles whose $R-v_R$ space distributions were presented on the right panel of Fig.~4 (CDM), as well as on the left (SIDM) and right (CDM+SNF) panels of Fig.~5 in \cite{2019MNRAS.485.1008B}. In these toy model simulations, SNF was implemented as a simplified analytical model and $\sigma_T/m_\chi = 2\,{\rm cm^2g^{-1}}$ was adopted as the fiducial SIDM cross section. 
    Below, we show the distributions of energy per unit mass of stars formed in three of the 
    hydrodynamic simulations introduced above. On the bottom left panel, we show the energy distribution of the 150 most metal-rich stars which are $0.8-0.9\,$Gyr old in the 
    smooth SF ($n_{\rm th} = 0.1\,{\rm cm^{-3}}$) CDM simulation (blue line) and the smooth SF SIDM simulation 
    ($\sigma_T/m_\chi = 1\,{\rm cm^2g^{-1}}$, red line). On the bottom right panel, we show the energy distribution of the 150 most metal-rich stars which are $0.8-0.9\,$Gyr old in the CDM simulations with smooth and bursty SF as blue (the same blue line as in the middle panel) and green lines, respectively. 
    Qualitatively, we find that the expectations of \cite{2019MNRAS.485.1008B} are verified: adiabatic energy injection (smooth SF) or redistribution (SIDM) results in a narrower energy distribution of kinematic tracers/stars compared to impulsive energy injection (bursty SF).}
    \label{fig:hists}
\end{figure*}

\begin{figure*}
    \centering
    \includegraphics[width=8cm,trim={0.5cm 0.5cm 0.5cm 0.5cm}, clip=true]{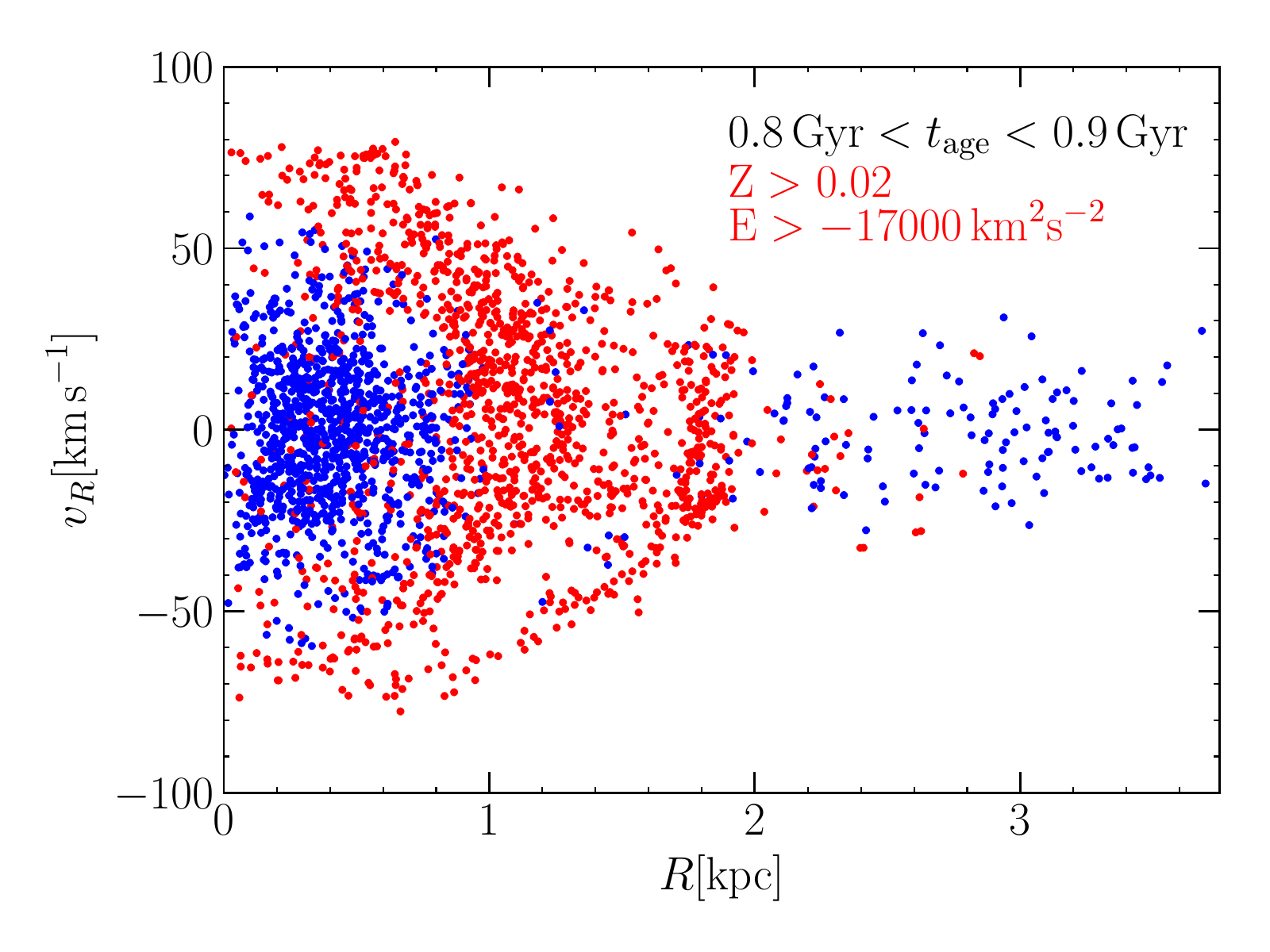}
    \includegraphics[width=8cm,trim={0.5cm 0.5cm 0.5cm 0.5cm}, clip=true]{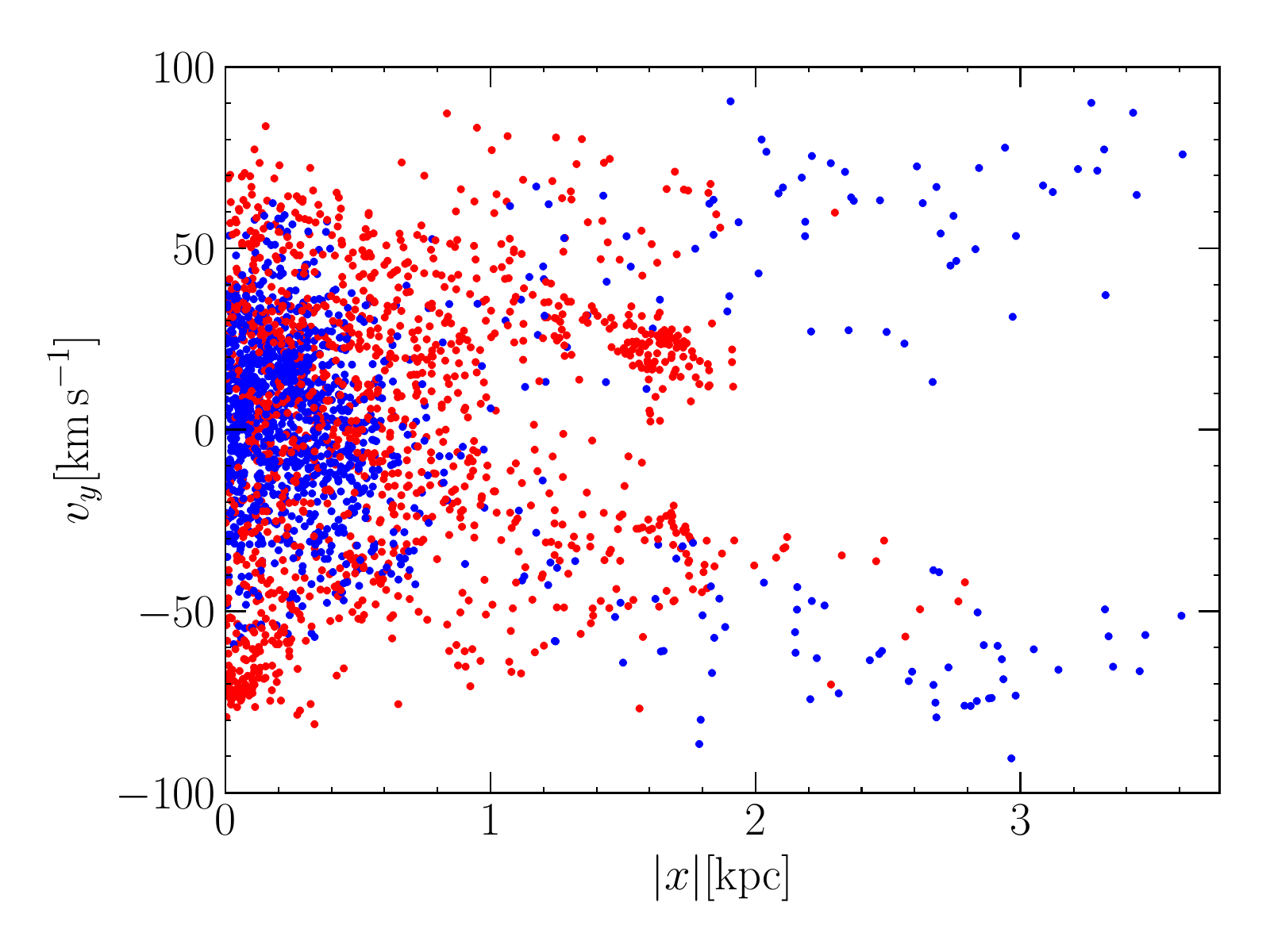}
    \caption{Shell-like signatures of early stage phase mixing caused by impulsive SN feedback in two different projections. We show scatter plots of the phase space coordinates of stars with ages in the range $0.8-0.9\,{\rm Gyr}$ after $3\,{\rm Gyr}$ of simulation time in the CDM run with $n_{\rm th} = 100\,{\rm cm^{-3}}$. We show two different projections of the phase space coordinates for all stars: ($R,v_R$) (left panel) and ($|x|,v_y$) (right panel). The star particles selected by the cuts indicated in the left panel are marked red, other star particles blue. We chose the indicated metallicity and specific energy cuts to select the stars that make up the shell-like feature in radial phase space. Notice that the clumps seen in the edge-on projection consist of those same star particles. }
    \label{fig:tagged}
\end{figure*}

\begin{figure*}
    \centering
    \includegraphics[width=8cm,trim={0.5cm 0.5cm 0.5cm 0.5cm}, clip=true]{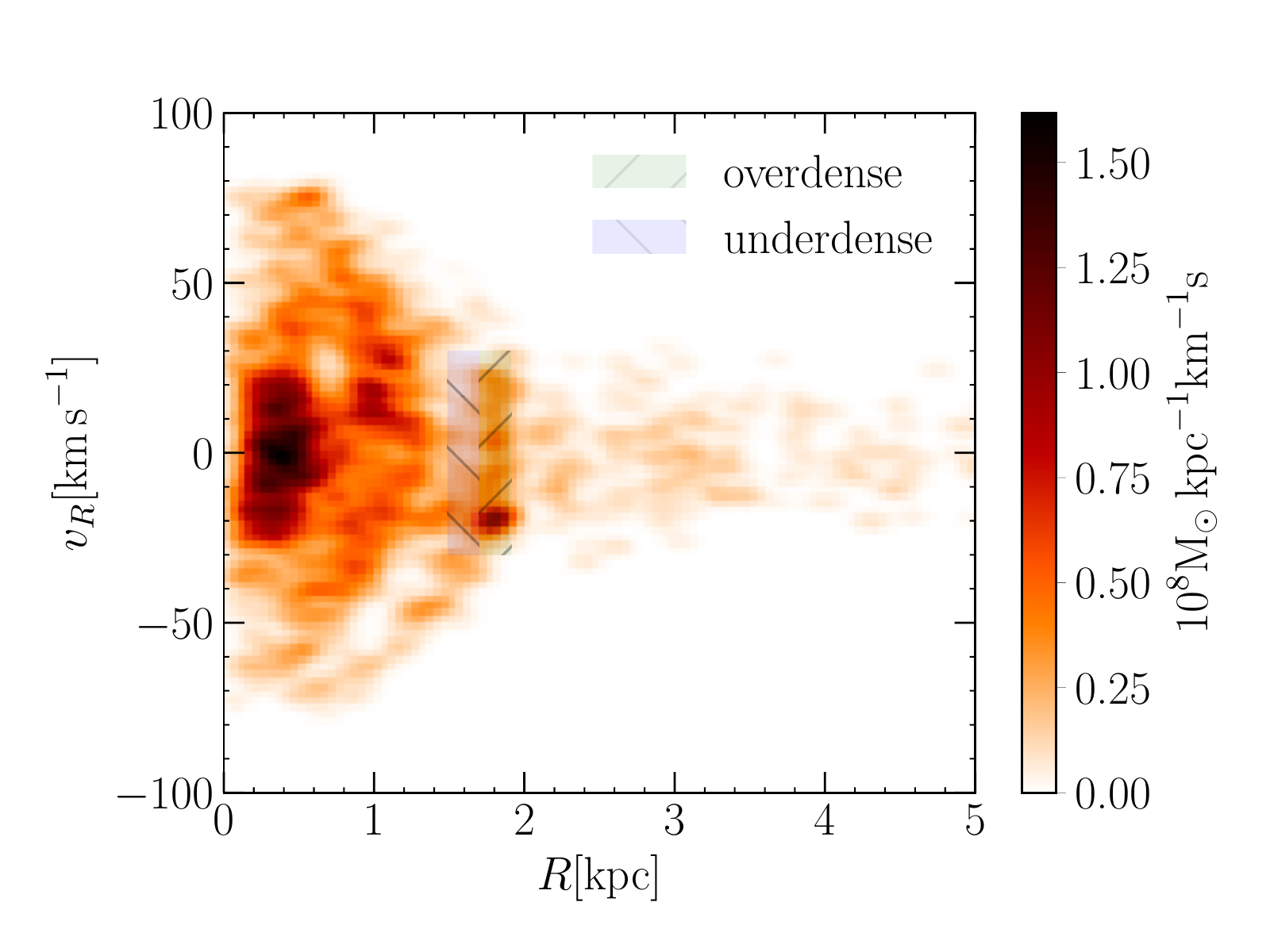}
    \includegraphics[width=8cm,trim={0.5cm 0.5cm 0.5cm 0.5cm}, clip=true]{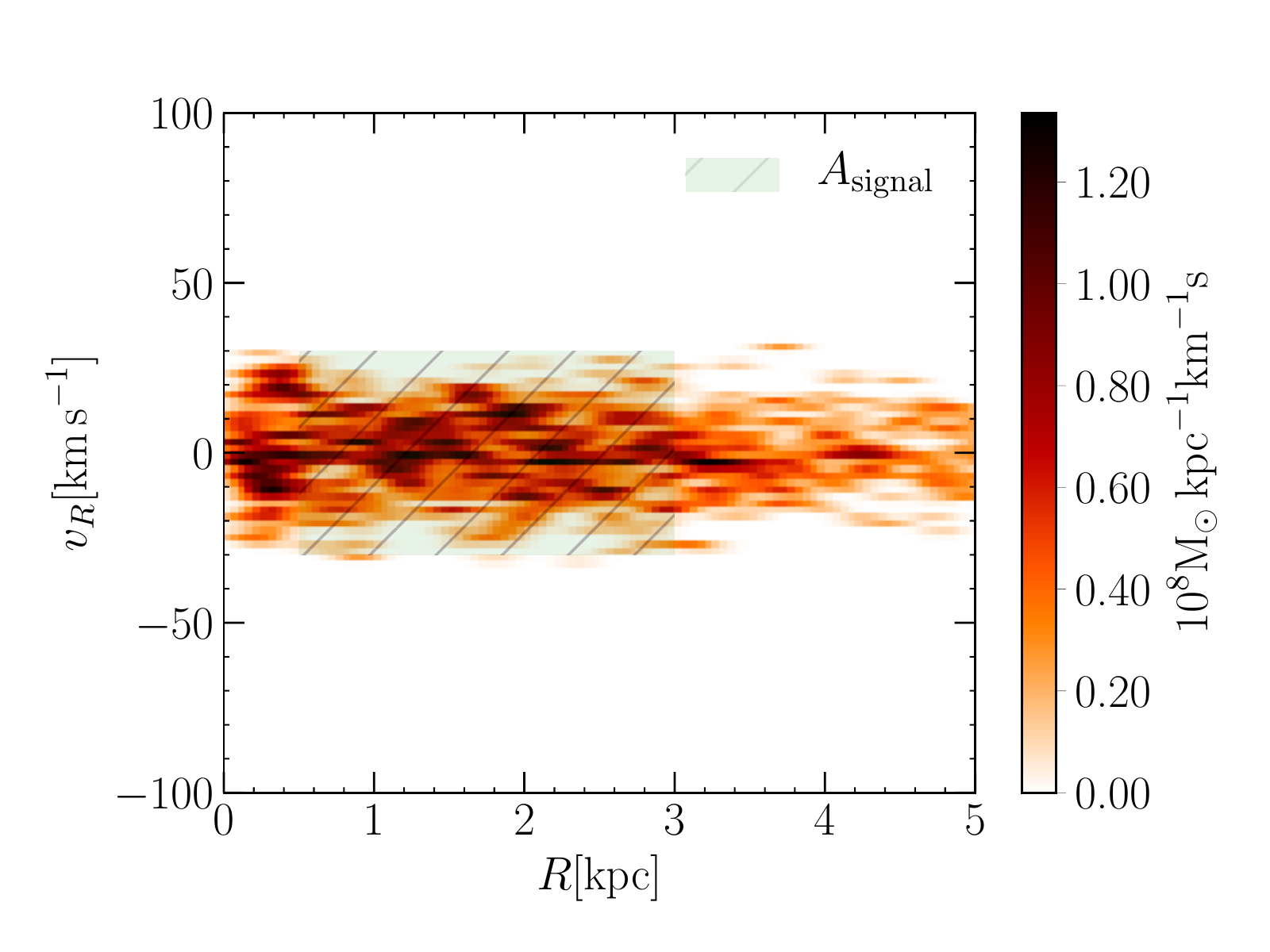}\\
    \includegraphics[width=0.6\linewidth]{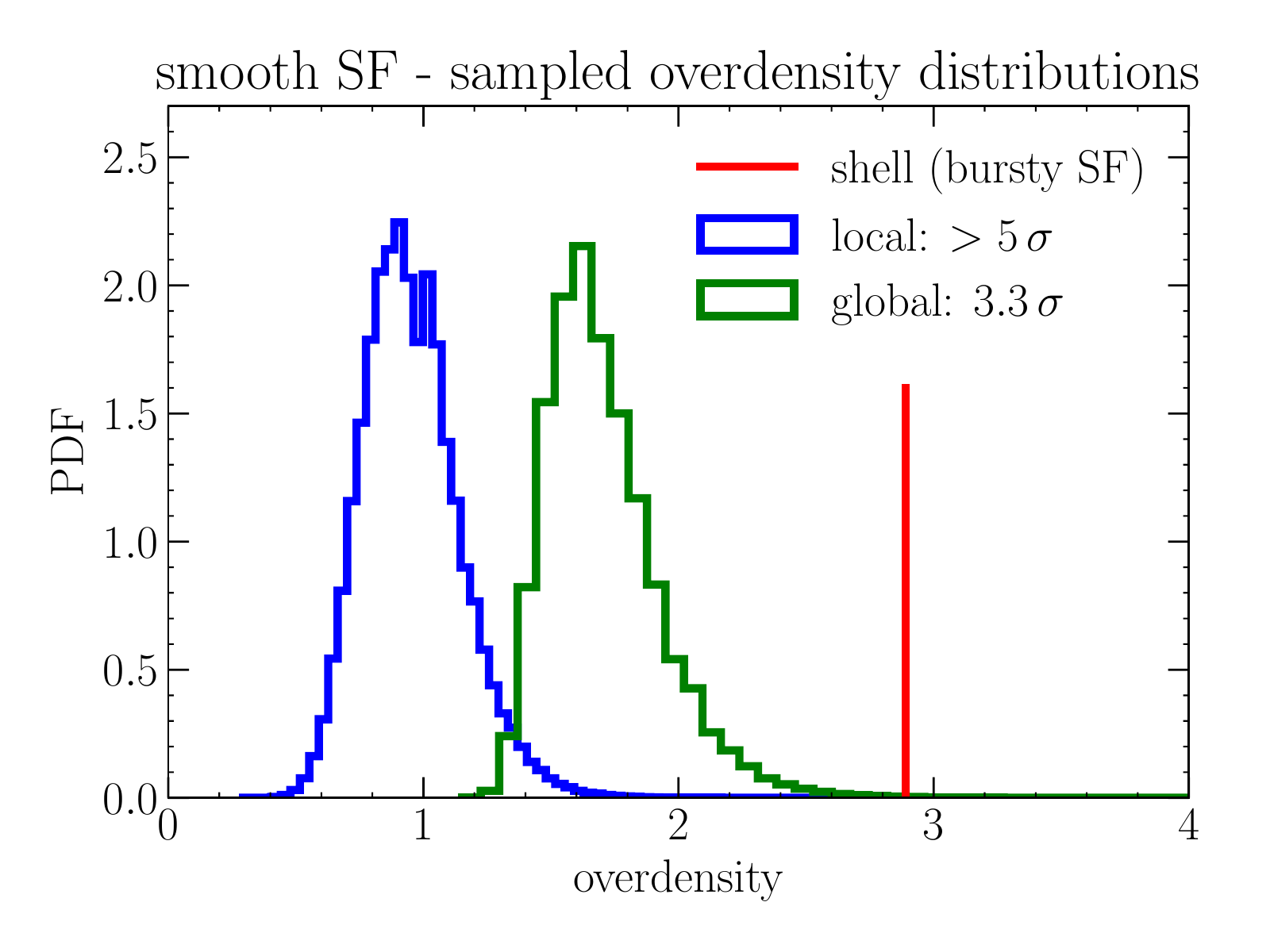}
    \caption{Significance of the overdense shell in the CDM simulation with bursty SF. The top left panel illustrates how the overdensity is defined. The green shaded rectangle defines the region in phase space in which the shell seen in the top left panel intersects the phase space region which is occupied in the smooth SF case (see top right panel; essentially defined by the range of radial velocities in the smooth SF case). Dividing the mass contained within this rectangular area in phase space by the mass contained within the adjacent blue shaded rectangle defines the shell overdensity. 
    The top right panel defines $A_{\rm signal}$, i.e. the area in phase space in which we expect shell-like overdensities. In the bottom panel we derive the significance of the shell-like overdensity as follows. Assuming that the underlying mass distribution is given by the smooth SF case, we use the mass distribution of the top right panel as a target distribution for random sampling. We resample the full distribution $10^7$ times, using the same number of particles as in the original distribution (top right panel). The resulting distribution of the local (global) overdensity is shown as a blue (green) curve. Here, local overdensity is defined using exactly the same phase space areas as in the top left panel, while global overdensity refers to the maximum overdensity over $A_{\rm signal}$ as defined in the top right panel. The red line is the measured shell overdensity. The estimated global (local) significance of the shell overdensity is $3.3\,\sigma$ ($>5\,\sigma$). }
    \label{fig:sig}
\end{figure*}

\begin{figure*}
    \centering
    \includegraphics[width=0.48\linewidth,trim={0.5cm 0.5cm 0.5cm 0.5cm},clip=true]{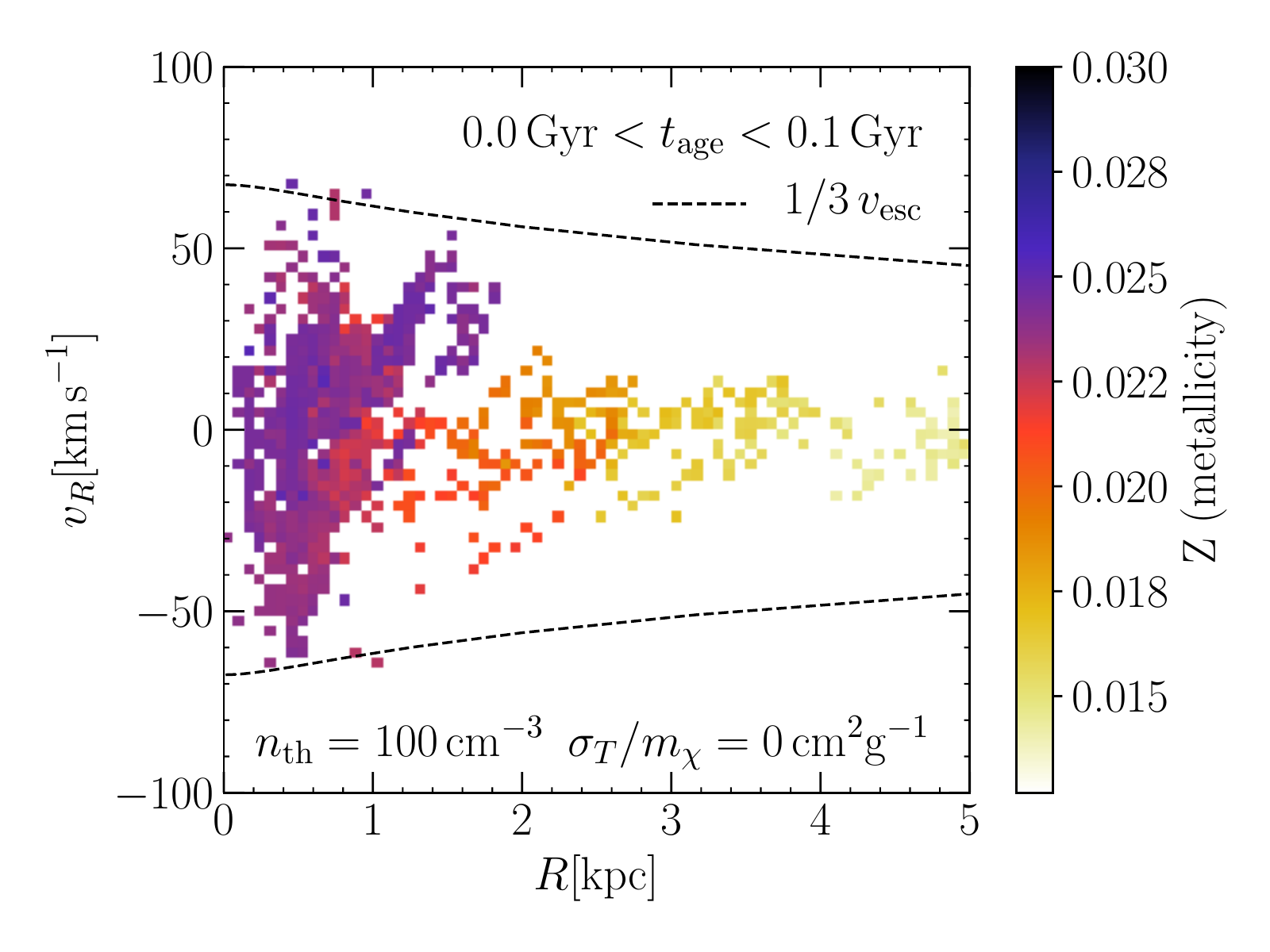}
    \includegraphics[width=0.48\linewidth,trim={0.5cm 0.5cm 0.5cm 0.5cm},clip=true]{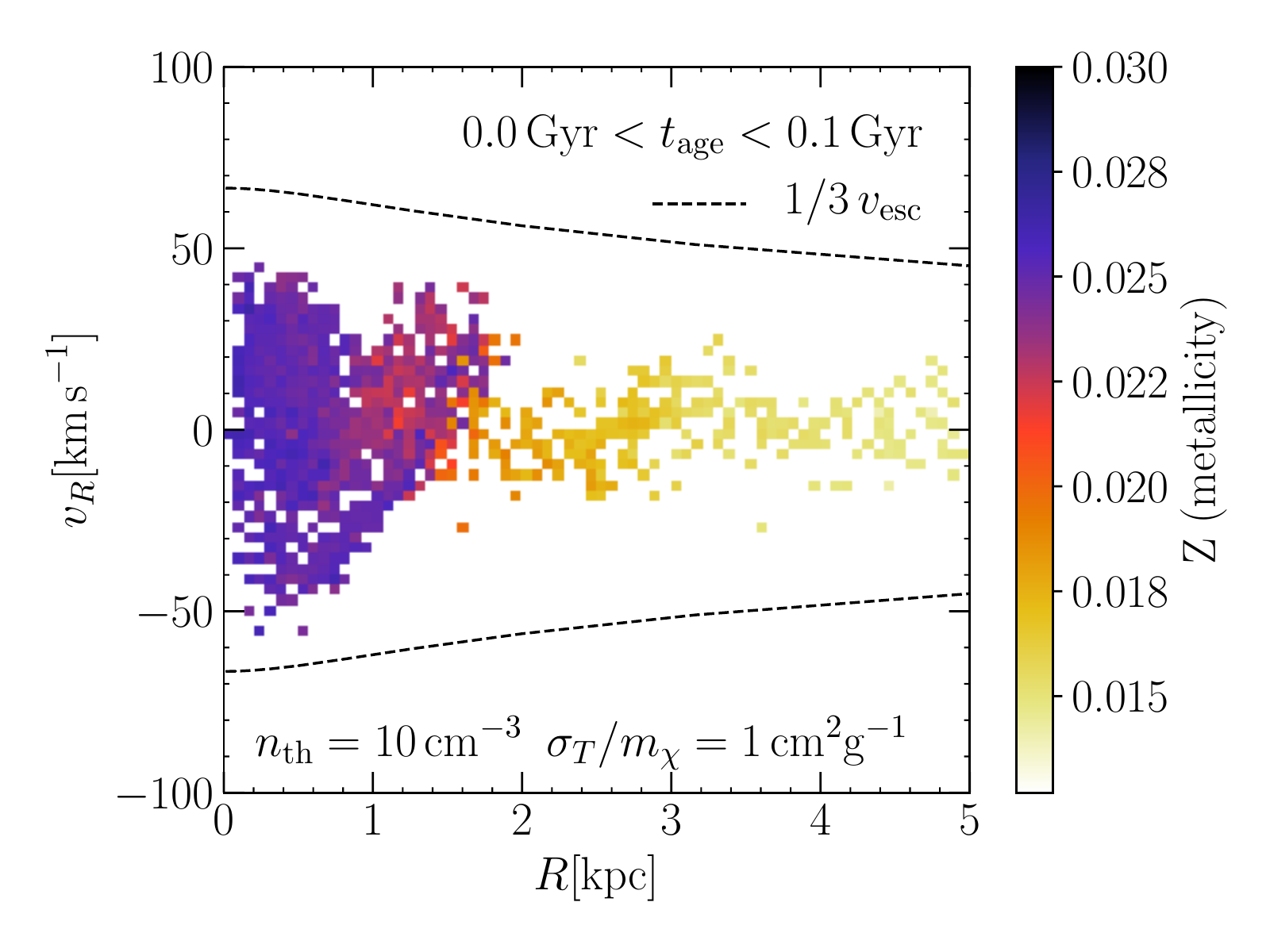}
    \caption{Left: Same as top right panel of Fig.~\ref{fig:features_main}, but after 2.2 Gyrs of simulated time instead of 3 Gyrs. Right: Same as top panel of Supplemental Fig.~\ref{fig:features_app}, but after 2.2 Gyrs of simulated time instead of 2.4 Gyrs.}
    \label{fig:early}
\end{figure*}

\begin{figure}
    \centering
    \includegraphics[width=\linewidth,trim={0.5cm 0.5cm 0.5cm 0.5cm},clip=true]{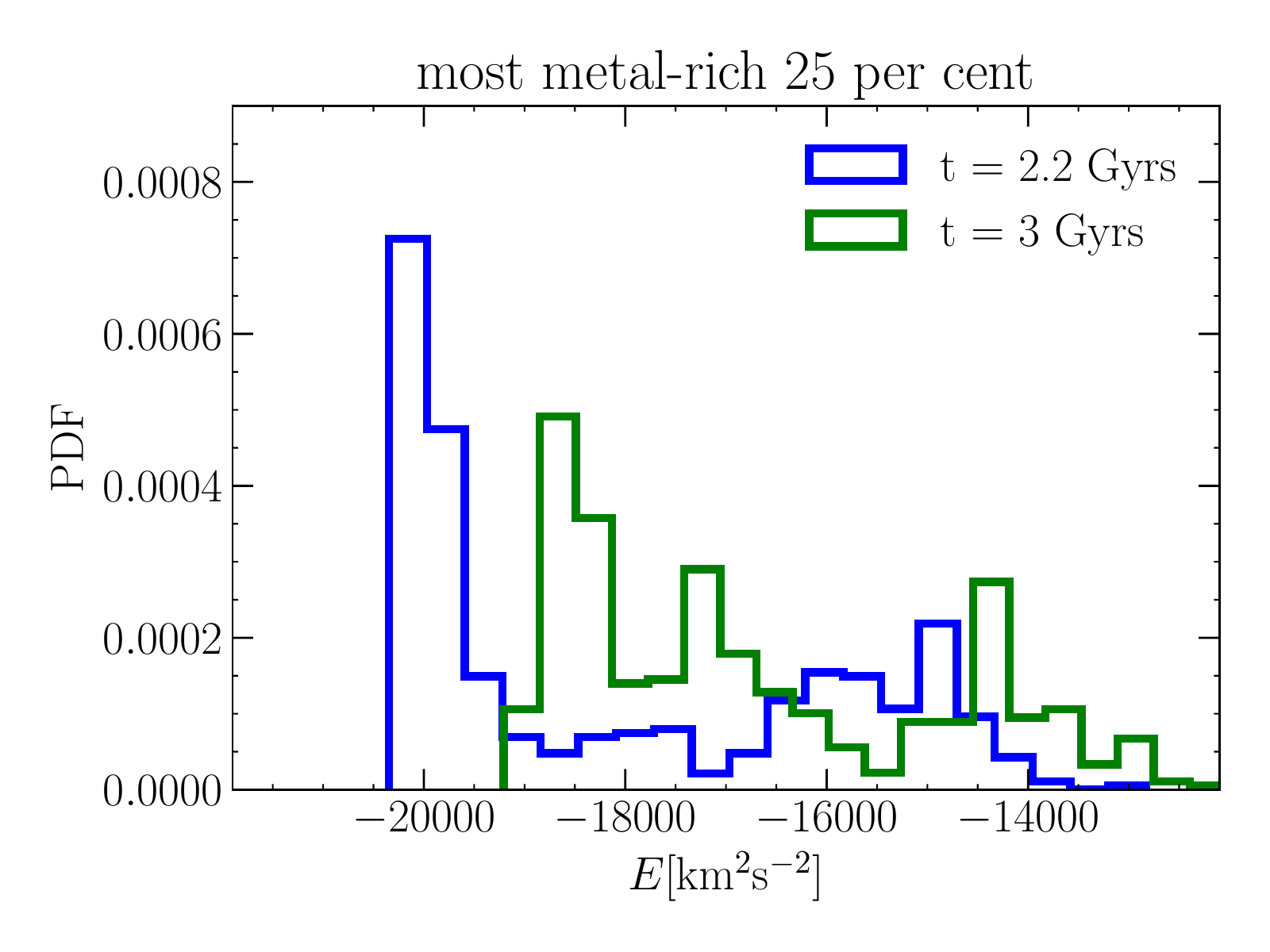}
    \caption{Energy distribution of the 25\% most metal-rich stars in the key phase space shell (right column of Fig.~\ref{fig:features_main}), compared to the energy distribution of the same stars shortly after they were born (at t=2.2 Gyrs). }
    \label{fig:edist_comp}
\end{figure}

\section{Supplemental Material: Computational methods}\label{sec:methods}

This {\it Letter} is based on a suite of 16 hydrodynamic simulations of an isolated dwarf galaxy in a live DM halo. Here we briefly outline how we set up the initial conditions, run the simulations, and analyze the 
results.

\subsection{Initial conditions}\label{supp_a}

We generate initial conditions of a self-gravitating 
system of a similar scale and global properties as the 
Small Magellanic Cloud \cite{2012MNRAS.421.3488H} at redshift zero with the important distinction that our simulated system is isolated. The initial conditions consist of four separate mass components: a spherically-symmetric DM halo with a Hernquist \cite{Hernquist:1990be} density profile, an exponential gas disc, an exponential ``stellar'' disc consisting of collisionless particles, and a ``stellar'' bulge following a Hernquist profile, also consisting of collisionless particles. We fix the structural parameters of the halo through its circular velocity at the virial radius $r_{200}$ and its NFW-equivalent concentration parameter $c_{200}$ \footnote{$r_{200}$ is defined through the virial mass of the equivalent NFW Halo, $M_{200} = 200\times 4\pi/3\rho_{\rm crit}r_{200}^3$ with $\rho_{\rm crit}$ the critical density of the Universe today, and $c_{200} = r_{200}/r_s$, with $r_s$ the equivalent NFW profile's scale radius. The scale parameter of a Hernquist halo is $a=r_s\sqrt{2[\log{c_{200}}-c_{200}/(1+c_{200})]}$ (see also \cite{2005MNRAS.361..776S}) and the total mass of the Hernquist sphere is fixed to $M_{200}$, which implies $M_{200,{\rm HG}} = r_{200}^2/(a+r_{200})^2{\rm M}_{200}$.}. In all our simulations, 
$v_{200} = 36.31\,{\rm km\,s^{-1}}$ and $c_{200} = 18$ initially. The corresponding virial mass of the system is $M_{200} \sim 1.6\times 10^{10}{\rm M}_\odot$ and we note that total halo mass, virial radius, and scale radius are fully determined by specifying $v_{200}$ and $c_{200}$. The gas disc is defined by its surface density profile,
\begin{align}
    \Sigma(R) = \frac{M_\infty}{2\pi H_{\rm gas}^2}\exp\left(-\frac{R}{H_{\rm gas}}\right),\label{eq:gas_surf}
\end{align}
where $M_\infty$ is the total mass of gas in the system, $H_{\rm gas}$ is a scale length, and $R$ is the polar radius in a coordinate system whose origin is at the centre of the galaxy and whose vertical axis is normal to the galactic plane. The scale length of the gas disc is $H_{\rm gas} = 2.1\,$kpc. $M_\infty$ is fixed as follows: we assume that the gas and stellar discs  
combined contain $4.45\%$ of the total mass of the dynamical system and that gas makes up for $84\%$ of the disc mass. In turn, this also fixes the mass of the stellar disc. The vertical structure of the gas disc is fixed by demanding that the gas is initially in hydrostatic equilibrium. We assume that the pressure of the gas is given by the equation of state of an isothermal ideal gas with a temperature of $T = 10^4$ K and that initially, the gas mixture has 
solar metallicity, $Z_\odot = 0.0127$ \cite{2009ARA&A..47..481A}, and is fully ionized. 
Under these conditions, and at a fixed polar radius $R$, the vertical structure of a gas disc in hydrostatic equilibrium is determined by the following set of equations:
\begin{align}
\frac{d\rho}{dz} &= -\frac{d\Phi}{dz}\frac{\rho^2}{P\Gamma}\label{eq:drdz}\\
\frac{d\Sigma}{dz} &= \rho.\label{eq:dsdz}
\end{align}
where $\rho$ is the gas density, $\Phi$ is the total gravitational potential, $P$ is the gas pressure, $\Gamma$ is the ratio of specific heats, and 
$\Sigma = \Sigma(z)$ denotes the cumulative gas surface density.
Using an initial guess for $\rho(R,0)$, we integrate equations \ref{eq:drdz} and \ref{eq:dsdz} upwards from the midplane and calculate $\rho(R,z)$ and $\Sigma(R,z)$ on a grid of $(R,z)$ values. The initially guessed value for $\rho(R,0)$ is multiplied by a correction factor and the vertical integration is repeated. The applied correction factor is the ratio between the known surface density (equation \ref{eq:gas_surf}) and the cumulative surface density at the largest $z$ value on the grid, $\Sigma(R,z_{\rm max})$, as calculated from equation \ref{eq:dsdz}. This vertical integration procedure is iterated five times.   
The surface density profile of the stellar disc has the same functional form as that of the 
gas disc, i.e., equation \ref{eq:gas_surf}, but with a scale length of $H_\ast = 0.7\,$ kpc. The star particles in the disc follow a ${\rm sech}^2$-distribution in the vertical direction,
\begin{align}
    \rho(R,z) \propto \Sigma(R)\left[\cosh\left(\frac{z}{z_0}\right)\right]^{-2},
\end{align}
with a vertical scale height $z_0 = 0.14\,$kpc. 
The star particles that make up the bulge are distributed following a Hernquist profile with scale length $A = 0.233\,{\rm kpc}$. 

Since both the DM halo and the bulge are spherically symmetric, we can use Eddington sampling to calculate their full distribution functions \cite{Eddington1916}. Note, however, that due to the presence of the baryonic discs, the total gravitational potential of the system, which appears in Eddington's integral, is not fully spherically symmetric. Therefore, we have to chose a preferred axis along which we evaluate the integral. Here,
we chose to evaluate Eddington's integral along the vertical direction and neglect the inaccuracy that is introduced by assuming isotropic distribution functions\footnote{We later let the system relax into a steady state. Thus, it is not crucial that the dynamical equilibrium is perfect at this stage.}. Having calculated the distribution functions, we assign velocities to the halo and bulge particles using a rejection sampling scheme. 

The velocities of the disc particles are calculated in a different way. First we calculate the streaming velocity and the velocity dispersion tensor on a logarithmic grid of radial and vertical coordinates, using the Jeans equation in cylindrical coordinates and the epicyclic approximation \cite{2005MNRAS.361..776S, 1993ApJS...86..389H}. Individual disc particles are then assigned velocities which are a sum of the streaming velocity at the position of the particle and a random component sampled from a local Maxwellian velocity distribution.
The motion of gas cells is given by the local streaming velocity of the gas. The streaming velocity field of the gas is calculated taking into account both gravity and pressure gradient \cite{2005MNRAS.361..776S}. 

Initially, our system consists of 
$1.2\times10^7$ DM particles, 
$4\times10^5$ gas cells, 
$8\times10^4$ disc particles, and 
$8\times10^3$ bulge particles. Each particle has an approximate mass of $1.3\times 10^3{\rm M}_\odot$. 
In a subsequent step, the system is enclosed within a cubic volume with a side length of $100\,{\rm kpc}$, into which a grid of background gas cells with low density and high temperature is introduced \cite{Springel:2009aa}. In the same step, previously constructed gas cells can be refined or de-refined if they contain a mass which is either larger than twice the average mass of a gas cell or smaller than half the average mass of a gas cell before introducing the background grid. 
Finally, since the resulting dynamical equilibrium is incomplete, we evolve the system for $1\,$Gyr in a simulation in which gas cooling and star formation are turned off. We adopt the final output of this preparation run as the relaxed initial conditions for our simulation suite.

\subsection{Simulations}\label{supp_b} 
We run all our 16 simulations using the 
code \texttt{AREPO} \cite{Springel:2009aa} which uses a TreePM algorithm to calculate the gravitational forces, 
while the hydrodynamics is 
solved on a moving mesh, using a quasi-Lagrangian scheme based on a Voronoi tessellation. 
We use the \texttt{SMUGGLE} \cite{2019MNRAS.489.4233M} stellar feedback model to model gas cooling, star formation, stellar evolution, supernova feedback, and radiative feedback. Self-interactions between DM particles are implemented using the Monte Carlo approach described in \cite{Vogelsberger2012}. 
Both \texttt{SMUGGLE} and the SIDM modules incorporated in \texttt{AREPO} 
can be calibrated through a number of parameters. Apart from the star formation threshold, all \texttt{SMUGGLE} model parameters are fixed to the values reported in table 3 of the original code paper \cite{2019MNRAS.489.4233M}. For the DM self-interactions, we adopt a fixed velocity-independent transfer cross section that varies between simulations. The amount of particles included in the nearest neighbour search for a scattering partner is set to $N_{\rm ngb} = 36\pm 5$. The model parameters that determine whether our modeled DM halo forms a constant density core or not are the SF threshold $n_{\rm th}$ and the self-interaction cross section $\sigma_T/m_\chi$.

In \texttt{SMUGGLE}, gas cells are converted into star particles following a standard probabilistic approach \cite{2003MNRAS.339..289S}. The local star formation rate in a given gas cell is 
\begin{align}
    \dot{M}_\star = \left\{ \begin{array}{ll}
        0\qquad &\rho < \rho_{\rm th} \\
        \epsilon \frac{M_{\rm gas}}{t_{\rm dyn}}\qquad &\rho \ge \rho_{\rm th}
    \end{array}  \right.,\label{eq:sfr}
\end{align}
where $\epsilon$ is an efficiency parameter which is set to $0.01$ in line with observations of slow SF in dense gas \cite{2007ApJ...654..304K}, $M_{\rm gas}$ is the gas mass in the cell and 
\begin{align}
    t_{\rm dyn} = \sqrt{\frac{3\pi}{32 G \rho_{\rm gas}}}
\end{align}
is the free-fall time of the gas. Stars can only form in gas cells in which the gas density is larger than $\rho_{\rm th}$, which is calculated from the numerical number density threshold $n_{\rm th}$ and the mean molecular weight of the gas. 
Moreover, large densities are reached faster in regions in which the potential well is deeper, i.e., towards the centre of galaxies. For larger SF thresholds, the gas must accumulate until it reaches the density threshold to become eligible for star formation,
leading to more concentrated and more bursty SF, both of which increases the efficiency of SNF at transforming DM cusps into cores \cite{2021ApJ...921..126B}.

In the stochastic SIDM model adopted here, the probability that a simulation particle {\it i} scatters with one of its nearest neighbours {\it j} in a given timestep is proportional to $\sigma_T/m_\chi$. Thus, the scattering rate increases if larger transfer cross sections are adopted. An increased scattering rate increases the rate at which the inner DM halo thermalizes, and thus 
reduces the time it takes to form the DM core. There is a relatively narrow window for the value of $\sigma_T/m_\chi$ to lead to fully isothermal cores in halos today. If the cross section is too small, $\sigma_T/m_\chi<0.1{\rm cm^2g^{-1}}$, only the innermost regions of the halo thermalize leading to a halo structure that is nearly indistinguishable from CDM \cite{Zavala2013,Rocha2013}.
On the other hand, adopting very large cross sections, $\sigma_T/m_\chi\gtrsim10{\rm cm^2g^{-1}}$, will trigger the gravothermal collapse phase of SIDM \cite{2002ApJ...568..475B,Colin2002,Koda2011,2015ApJ...804..131P,2020PhRvD.101f3009N}, which eventually results in SIDM halos that are even cuspier than CDM halos. 

In our simulation suite, we adopt $4\times 4$ combinations of $n_{\rm th}$ and $\sigma_T/m_\chi$. The SF threshold takes values of $n_{\rm th} = (0.1,1,10,100)\,{\rm cm^{-3}}$, whereas for the SIDM transfer cross section we adopt values of $\sigma_T/m_\chi = (0,0.1,1,10)\,{\rm cm^2g^{-1}}$. 

In all our production runs we adopt a softening length $\epsilon = 24\,$pc for all particle species. In order to properly resolve the effects of supernova feedback in runs with large SF thresholds \cite{2020MNRAS.499.2648D}, this softening length is slightly smaller than the optimal softening length for a system of this size \cite{Power:2002sw}.

\subsection{Post processing}\label{supp_c}

The gas density projections shown in Fig.~\ref{fig:galaxyview} and the stellar light projections shown in Supplemental Fig.~\ref{fig:galaxyview_supp} are created as in \cite{2019MNRAS.489.4233M}. The density projection panels are obtained by integrating for each pixel of the image the total gas density along the line of sight for a depth equal to the image side-length. The stellar light images are generated using stellar population synthesis models coupled to a line-of-sight dust extinction calculation assuming a constant dust-to-metals ratio.

Here we briefly discuss how we process simulation snapshots to create the remaining figures. We start by calculating the 
centre of potential $\mathbf{R}$ 
and velocity $\mathbf{V}$ of the DM system, using a shrinking spheres method. A first guess is obtained by summing over all $N$ DM particles,
\begin{align}
    \mathbf{R}^{0} &= \frac{\sum_{i=1}^{N_{\rm DM}}\Phi(\mathbf{r}_i)\,\mathbf{r}_i}{\sum_{i=1}^{N_{\rm DM}}\Phi(\mathbf{r}_i)}\\
    \mathbf{V}^{0} &= \frac{\sum_{i=1}^{N_{\rm DM}}\Phi(\mathbf{r}_i)\,\mathbf{v}_i}{\sum_{i=1}^{N_{\rm DM}}\Phi(\mathbf{r}_i)},
\end{align}
where $\mathbf{r}_i$ and $\mathbf{v}_i$ are the position and velocity vectors of particle {\it i} and $\Phi$ is the gravitational potential.  
In subsequent steps, the sums are restricted to DM particles that are closer to the currently estimated centre of potential than a given threshold radius $r_{\rm th}$. For instance, in step $n+1$, particles need to satisfy $|\mathbf{r}-\mathbf{R}^n| < r_{{\rm th},n+1}$ to be included in the sum. Here we use three iterations, with threshold radii of $50\,$kpc, $5\,$kpc, and $0.5\,$kpc. Once the final values of $\mathbf{R}$ and $\mathbf{V}$ have been calculated, we shift the phase space coordinates of all particles and gas cells, defining a new coordinate system in which the centre of potential of the system is at $(0,0,0)$ and the system has no bulk motion. 

Subsequently, since we simulate the evolution of a rotationally supported galaxy, it is convenient to perform a coordinate transformation such that the net rotation occurs in the $x-y$ plane of the new coordinate system. To that end, we calculate the total angular momentum from all particles which are part of the rotating disc, i.e., the collisionless disc particles and the formed star particles, as well as the gas cells. Since individual simulation particles represent stellar populations and gas cells represent an extended volume of moving gas, we have to give a different weight to the angular momenta of particles/gas-cells with different masses. Hence, we calculate the total angular momentum as a mass-weighted sum of the angular momenta of all rotating particles and gas cells. We then rotate the coordinates and velocities of all particles into a coordinate system in which the vertical axis is aligned with the direction of the total angular momentum vector. All figures presented here are constructed from dynamical quantities measured in this coordinate system.

The density and velocity dispersion profiles shown in Fig.~\ref{fig:dm_profiles} and Supplemental Fig.~\ref{fig:dm_profiles_supp} are calculated in 20 bins which are equally spaced in logarithmic radius. Fig.~\ref{fig:features_main} and Supplemental Figs.~\ref{fig:features_projected} and \ref{fig:features_app} are subject to the age cuts stated in their respective captions. The phase space binning adopted in Fig.~\ref{fig:features_main} and Supplemental Fig.~\ref{fig:features_app} is as follows. We focus on the radial range $(0\,{\rm kpc} < R < 5\,{\rm kpc})$ and 
the radial velocity range $(-100\,{\rm km\,s^{-1}} < v_R < 100\,{\rm km\,s^{-1}})$. 
To produce the two-dimensional colour plots shown in the top row of Fig.~\ref{fig:features_main} and in Supplemental Fig.~\ref{fig:features_app}, we divide the $R-v_R$ plane into $70\times 70$ bins that equidistantly cover the adopted ranges of radii and radial velocities. The metallicity values shown in Fig.~\ref{fig:features_main} and Supplemental Fig.~\ref{fig:features_app} are mass-weighted averages over all stellar particles within a given bin. The two-dimensional colour plots in the top row of Supplemental Fig.~\ref{fig:features_projected} are constructed in the same way, using the phase space coordinates $|x|$ and $v_y$ instead of $R$ and $v_R$. To construct the two-dimensional colour plots shown in the bottom rows of Fig.~\ref{fig:features_main} and Supplemental Fig.~\ref{fig:features_projected} we have used the scipy routine {\it stats.gaussian\_kde} with a fixed scalar bandwidth of $0.07$. The escape velocities shown in Fig.~\ref{fig:features_main} and Supplemental Fig.~\ref{fig:features_app} are calculated from the gravitational potential as 
\begin{align}
   v_{\rm esc} = \sqrt{-2\Phi(r)}, 
\end{align}
where the gravitational potential is calculated in the same 20 bins as the density profile and the velocity dispersion profile. 

The energy distributions displayed in the bottom panels of Supplemental Fig.~\ref{fig:hists} are obtained from the 150 most metal-rich stellar particles of a given age in three different simulations as stated in the caption. The energy of each stellar particle is comprised of its kinetic energy, which we calculate using the shifted velocities, and its potential energy, which is calculated by \texttt{AREPO} on the fly.  
In the top panel of Supplemental Fig.~\ref{fig:hists} we show the energy distribution of {\it orbital families} of kinematic tracers in a DM halo that has formed a core either adiabatically or impulsively 
simulated using a spherically symmetric toy model \cite{2019MNRAS.485.1008B}. Similar to the energies of the stellar particles formed in the hydrodynamic simulations presented here, the energies of the tracers are calculated in a shifted coordinate system in which the centre of potential is at $(0,0,0)$ and the system has no bulk motion. However, since the simulated toy model systems are spherically symmetric and isotropic, there is no net angular momentum and hence we do not need to rotate into a different coordinate frame. 
Finally, we note that the SF histories shown in Supplemental Fig.~\ref{fig:sfr} are calculated by \texttt{SMUGGLE} on the fly and that Supplemental Fig.~\ref{fig:tagged} is created by simply plotting the phase space coordinates of all stars of the indicated age and that all stars that fulfill the cuts stated in the left panel are coloured in red.

\subsection{Determining the significance of the shell overdensity}\label{supp_d}
We aim to obtain a conservative estimate for the significance of the overdense shell in the CDM run with bursty SF (see right column of Fig.~\ref{fig:features_main}). To that end, we define the shell overdensity as the ratio between the mass contained within the rectangle in phase space defined by [$1.7 < R/{\rm kpc} < 1.9125, -30 < v_R/({\rm km\,s^{-1}}) < 30$] and the mass contained within the rectangle [$1.4875 < R/{\rm kpc} < 1.7, -30 < v_R/({\rm km\,s^{-1}}) < 30$] and find that the overdensity is $\Delta_{\rm shell} = 2.89$ (see Supplemental Fig.~\ref{fig:sig}). We now look to determine the likelihood for such an overdensity to arise as a random fluctuation in the smooth SF distribution. To that end, we take the normalized cumulative radial mass distribution, $\bar{M}(R)$, of the CDM run with smooth SF (left column of Fig.~\ref{fig:features_main}) as a target distribution for random sampling. We then re-sample this distribution a total of $10^7$ times, using the same total number of stars as in the original (simulated) distribution. For each sampled distribution, we calculate the density ratio between the same two rectangles as in the bursty SF case, defining the ``local'' overdensity. We also look for ``global'' overdensities, i.e., the largest overdensities within a signal region, $A_{\rm signal}$, which is defined by [$0.5 < R/{\rm kpc} < 3, -30 < v_R/({\rm km\,s^{-1}}) < 30$] (see top right panel of Supplemental Fig.~\ref{fig:sig}) and corresponds to the phase space region in which we found shell-like overdensities across all our runs with bursty SF. We look for overdensities in this region by calculating the ratios between adjacent phase space rectangles of the same size as above, and covering the same range of radial velocities. We scan the radial range of $A_{\rm signal}$ in steps which are equal to the softening length adopted in the simulations, i.e., the resolution limit. In the bottom panel of Supplemental Fig.~\ref{fig:sig}, we present the resulting distribution of local (global) overdensities as a blue (green) line. We also show the shell overdensity in red. We calculate the local (global) significance from the number of sampled distributions in which the local (global) overdensity $\Delta < \Delta_{\rm shell}$ by assuming that the local (global) overdensities follow a Gaussian distribution. To determine the minimum number that is required for the shell to have a global significance which is $> 2\sigma$ we repeat the re-sampling procedure several times, each time with a smaller number of (sampled) stars. 

\section{Supplemental Material: Shell formation}\label{sec:shell_formation}

In this section, we aim to provide further insight into the proposed mechanism for the creation of phase space shells, which we outline in Section \ref{main_b}. The foundation of our proposed mechanism is twofold. The first part relies on results of \cite{2016ApJ...820..131E}, who showed that stars that are born in a turbulent ISM inherit the orbits of the clouds of star-forming gas, and that their orbits are then quickly heated further by subsequent episodes of impulsive SNF.

The second part relies on the findings of \cite{Pontzen:2011ty}, in particular on the fact that the energy change that a particle experiences in response to an impulsive change in the gravitational potential is a function of that particle's orbital phase at the time at which this change occurs.
As shown in \cite{2019MNRAS.485.1008B} and \cite{2021ApJ...921..126B}, this implies that repeated impulsive feedback widens the energy distribution of orbital families, and can lead to the formation of shells such as the ones reported here.

In order for this two-step process to be a self-consistent explanation for the formation of shells, the stars that are formed in starburst events (and thus have similar metallicities and formation times) need to form orbital families -- implying that their orbital phases should be similar at the time they are born. While this seems like a reasonable assumption, we can compare the shells reported here to the phase space distribution of the same stars at earlier times to see if this assumption does indeed hold true.

Supplemental Fig.~\ref{fig:early} shows the $R-v_R$ space distributions of the same stars as in the top right panel of Fig.~\ref{fig:features_main} (on the left panel) and the top panel of Supplemental Fig.~\ref{fig:features_app} (on the right panel).
As we can see, no shells are present yet in either of the two panels. On the right panel, we see that a large number of recently born high metallicity stars are on an infalling orbit, and that a contiguous narrow region in R-vR space is occupied. This is exactly as expected in the proposed mechanism. On the left panel, the situation is a bit more complex. Stars still occupy a contiguous region in R-vR space, suggesting that they were born together. However, the positive radial velocities that increase with radius suggest that they are -- at the time the snapshot was taken -- already being pushed outwards by supernovae, likely generated by stars born in this starburst or shortly before.

In both panels above, the range of orbital phases occupied by high-metallicity stars is significantly narrower than later on, when we observe fully formed shells. We take this as proof that stars which are born in one single starburst event in a turbulent ISM do indeed form an orbital family. This orbital family can then later be split by subsequent episodes of impulsive SNF.

According to our proposed formalism, this split is the consequence of a widening of the stars' energy distribution, which is caused by the impulsive feedback episodes. In order to verify this part of our proposed formalism, we compare the (normalized) energy distribution of the 25\% most metal rich stars in the top right panel of Fig.~\ref{fig:features_main} to the (normalized) energy distribution of those same stars shortly after they were born (after 2.2 Gyrs of simulated time, corresponding to the phase space distribution shown in Supplemental Fig.~\ref{fig:early}).
This comparison is shown in Supplemental Fig.~\ref{fig:edist_comp}. We find that the final energy distribution is significantly less peaked, and a broader range of energies is occupied. On average, energies have increased in magnitude as well, which makes sense, since core formation occurs due to this exact same increase in orbital energy, only in DM particles instead of stars. Throughout the simulation, this means that the overall orbital energies of metal-rich stars have increased, and that the energy distribution has been widened in response to impulsive feedback. Together, Supplemental Figs.~\ref{fig:early} and \ref{fig:edist_comp} confirm our proposed mechanism for the formation of phase space shells.

\begin{acknowledgments}
JB and JZ acknowledge support by a Grant of Excellence from the Icelandic Center for Research (Rann\'is; grant numbers 173929 and 206930). The simulations presented here were carried out
on the Garpur supercomputer, a joint project between the University of Iceland and University of Reykjav\'ik with funding from Rann\'is. LVS is grateful for financial support from NASA ATP
80NSSC20K0566, NSF AST 1817233 and NSF CAREER 1945310
grants. PT acknowledges support from NSF grant AST-1909933, AST-2008490, and NASA ATP Grant 80NSSC20K0502. FM acknowledges support through the program ``Rita Levi Montalcini'' of the Italian MUR. PT acknowledges support from National Science Foundation (NSF) grants AST-1909933, AST-200849 and National Aeronautics and Space Administration (NASA) Astrophysics Theory Program (ATP) grant 80NSSC20K0502.
\end{acknowledgments}

\bibliography{main}

\end{document}